\documentclass{article}
\usepackage{amsfonts}
\usepackage{amsmath, amssymb}
\usepackage{hyperref}
\hypersetup{hidelinks,pdftitle={Critical behavior of the bosonized Schwinger model with uniform matrix product states},pdfauthor={Takuya Okuda}}
\usepackage{graphicx}
\usepackage{braket}
\usepackage{bm}
\usepackage{booktabs}
\usepackage{placeins}
\usepackage{needspace}
 
\usepackage[margin=1in]{geometry}

\newcommand{\mcval}{0.333\,555}
\newcommand{\mcerr}{(5)}
\newcommand{\mcvalerr}{\mcval\mcerr}

\title{Critical behavior of the bosonized Schwinger model\\
 with uniform matrix product states}

\author{Takuya Okuda\thanks{takuya@hep1.c.u-tokyo.ac.jp}\\[0.5ex] 
\normalsize Graduate School of Arts and Sciences, University of Tokyo\\
\normalsize Komaba, Meguro-ku, Tokyo 153-8902, Japan}

\date{November 12, 2026}
\begin{document}
\maketitle

\begin{abstract}
Quantum electrodynamics in two spacetime dimensions, also known as the Schwinger model, admits a dual formulation in terms of a non-compact boson field.
Using a Hamiltonian lattice discretization of the bosonized model, we compute the critical mass of the continuum Schwinger model at $\theta=\pi$.
We numerically obtain the ground state on an infinite chain by the variational uniform matrix product state (VUMPS) algorithm, preserving charge conjugation exactly.
As a truncated basis for the local Hilbert space, we employ the oscillator basis that diagonalizes the one-site reduced density matrix of the free limit.
We constrain the fitting functions for the continuum extrapolation using the Symanzik effective theory for lattice corrections.
Our result for the critical mass, $(m/g)_c = 0.333\,555(5)$, agrees with recent staggered-fermion determinations at comparable precision.

\end{abstract}

\tableofcontents

\section{Introduction}
\label{sec:intro}

The Schwinger model is quantum electrodynamics in one space and one time dimension~\cite{Schwinger:1962tn,Schwinger:1962tp}.
It admits a topological term and the massive theory depends on its coefficient, the theta angle $\theta$.
At $\theta=\pi$ the theory undergoes a transition to a phase with spontaneously broken charge conjugation as the fermion mass increases.
This phase transition, predicted by Coleman~\cite{Coleman:1976uz}, belongs to the two-dimensional Ising universality class, as supported by early Hamiltonian calculations~\cite{Hamer:1982mx}, DMRG~\cite{Byrnes:2002nv}, and Grassmann tensor renormalization group calculations~\cite{Shimizu:2014fsa}.
The dimensionless critical mass $(m/g)_c$, where $g$ is the gauge coupling, provides a benchmark for Hamiltonian lattice calculations.
Gauge-invariant matrix product states (MPS) have been used to calculate ground-state and excitation properties of the Schwinger model and extrapolate them to the continuum~\cite{Buyens:2013yza}.
Recent determinations give
$0.333\,561(4)$ from the density-matrix renormalization group (DMRG) on finite lattices~\cite{ArguelloCruz:2024xzv}
and
$0.333\,556(5)$ from gauge-invariant variational uniform matrix product states (VUMPS)~\cite{Fujii:2024giv},
both in the staggered fermion formulation.
Their difference is comparable to each quoted uncertainty.
Since both calculations use staggered fermions, a different regularization offers a check with different lattice artifacts.

Hamiltonian simulations of gauge theories with bosonic degrees of freedom require control of both the local Hilbert-space truncation and the spatial lattice spacing.
Tensor-network calculations introduce a further approximation through the bond dimension.
These three dependences must be examined separately when comparing continuum predictions.
The local truncation and lattice-spacing dependence also matter for quantum simulations of a bosonic Hamiltonian, for which classical calculations can supply reference results~\cite{Klco:2018zqz,Calliari:2024ppe}.

Numerical studies of the bosonized lattice Schwinger model include the Monte Carlo calculation of the static potential by Bender, Rothe, and Rothe~\cite{Bender:1984qg}.
Ohata~\cite{Ohata:2023sqc} developed this approach further, treating lattice normal ordering explicitly and studying the dependence on $\theta$ without a sign problem.
A subsequent study~\cite{Ohata:2023gru} examined the finite-temperature phase diagram near the critical point at $\theta=\pi$ and found electric-field correlations consistent with Ising scaling.
These Monte Carlo calculations use a Euclidean spacetime lattice.

In this paper, we determine the critical mass using a bosonized lattice Hamiltonian.
Time remains continuous, and we calculate the ground state on an infinite spatial chain.
At zero fermion mass, this Hamiltonian follows from gauging a $U(1)$ symmetry of the compact boson in the modified Villain formulation~\cite{Fazza:2022fss,Seifnashri:2026ema}.
Solving the gauge constraints leaves a non-compact scalar whose quadratic potential comes from the electric energy.
We add the fermion mass term, which becomes a cosine interaction through continuum bosonization~\cite{Coleman:1975pw,Coleman:1976uz}, and solve the resulting chain with the VUMPS algorithm~\cite{ZaunerStauber:2018vumps}.
Gauging the fermionic Villain model~\cite{Lu:2026fvm} gives a Hamiltonian that differs by irrelevant terms, whose effects vanish in the continuum limit.

The same bosonized lattice Hamiltonian as ours and the VUMPS method were used in~\cite{Belyansky:2024scattering} to study quark and meson scattering, including inelastic particle production and string breaking, at fixed lattice spacings.
That work also proposes an analog implementation using circuit quantum electrodynamics (circuit QED), connecting this bosonic formulation to quantum simulation.
Here we investigate how its critical mass approaches the continuum value, with separate checks of the local-state truncation and the finite bond dimension.
Since the bosonic local Hilbert space has infinite dimensions, numerical simulation needs a cutoff on the number of states retained at each site.
We choose these states by an optimization in the free bosonic theory and evaluate the cosine interaction using exact harmonic-oscillator matrix elements.
The local cutoff limits the number of states at each site. 
The MPS bond dimension instead limits the entanglement between the two parts of the chain separated by a bond.
Charge conjugation~$C$ acts as $(-1)^n$ on the oscillator level~$n$.
Keeping this symmetry exact separates the eigenvalues of the MPS transfer matrix into even and odd sectors and identifies those that govern the correlation length of the $C$-breaking order parameter.

To guide the continuum extrapolation, we use the Symanzik effective theory~\cite{Symanzik:1983dc,Symanzik:1983gh,Weisz:2010rg}, which is a framework that represents lattice corrections by local operators added to the continuum action or Hamiltonian.
The coefficients of these operators can be determined perturbatively from correlation functions of the free massless non-compact boson.
Within this local expansion, charge counting and reflection symmetry allow only even powers of $ga$, with possible logarithmic factors.
We then assume that the location of the critical point has a Taylor expansion in the operator coefficients. Under this assumption, the critical mass inherits the same dependence on the lattice spacing.
We average the results from several fitting functions and lattice-spacing ranges using Akaike weights~\cite{Jay:2020jkz,Neil:2022joj}, and test how the result changes when we include higher-order logarithms or alter the averaging prescription.

Section~\ref{sec:formulation} presents the lattice Hamiltonian, the local oscillator basis, and the implementation of charge conjugation.
Section~\ref{sec:fes} explains how we locate the critical point and extrapolate in bond dimension.
Section~\ref{sec:results} gives the continuum fits, the resulting critical mass, and the uncertainty estimates.
Section~\ref{sec:c2} explains the physical argument for the lattice-spacing dependence and its assumptions.
We conclude in Sec.~\ref{sec:conclusion}.
Appendix~\ref{sec:freeappendix} derives the free-theory results used to construct the local basis, and Appendix~\ref{sec:tfim} tests the bond-dimension extrapolation in the transverse-field Ising chain.
Appendix~\ref{sec:freecorrelatormatching} gives the detailed correlation-function matching, including the normal-ordering conventions, local subtractions, and an example of determining two coefficients.

\section{Bosonized lattice formulation and optimized local basis}
\label{sec:formulation}

\subsection{Lattice Hamiltonian and normal ordering}

The continuum bosonized Schwinger Hamiltonian~\cite{Coleman:1975pw,Coleman:1976uz,Okuda:2022hsq} is
\begin{equation}
H_{\mathrm{cont}} = \int dx\left(
\tfrac{1}{2}\Pi_\phi^2 + \tfrac{1}{2}(\partial_x\phi)^2
+ \frac{g^2}{2\pi}\,\phi^2
- mg\,\frac{e^{\gamma}}{2\pi^{3/2}}
:\!\cos\bigl(2\sqrt{\pi}\,\phi-\theta\bigr)\!:_{\infty}
\right).
\label{eq:Hcont}
\end{equation}
Here $m$ is the fermion mass, $\phi$ the non-compact boson field, and $\Pi_\phi$ its canonical momentum density; $g$ is the gauge coupling and $\gamma$ is the Euler constant.
The notation $:\bullet:_\infty$ denotes normal ordering with respect to the infinite-volume vacuum at zero fermion mass. 
This free boson has mass $g/\sqrt{\pi}$. 
The expression $g e^{\gamma}/(2\pi^{3/2})$ is minus the value of the chiral condensate $\langle \bar\psi\psi\rangle$ in the massless theory at $\theta=0$, where $\psi$ is the fermion field of the Schwinger model, and $\bar\psi$ is its Dirac adjoint.
On an infinite chain of spacing~$a$, the canonical variables are $\phi_j=\phi(ja)$ and $p_j=a\Pi_\phi(ja)$, with $[\phi_j,p_k]=i\delta_{jk}$.
Both are dimensionless.
We use
\begin{equation}
H_{\mathrm{lat}} = \frac{1}{2a}\sum_j\Bigl(p_j^2
+ (\phi_{j+1}-\phi_j)^2\Bigr)
+ \frac{a g^2}{2\pi}\sum_j \phi_j^2
- mg\,\frac{e^{\gamma}}{2\pi^{3/2}}\,a\sum_j
:\!\cos\bigl(2\sqrt{\pi}\,\phi_j-\theta\bigr)\!:_{\infty}.
\label{eq:Hlat}
\end{equation}
At $m=0$, this Hamiltonian follows from gauging a compact boson theory in the modified Villain formulation~\cite{Fazza:2022fss,Seifnashri:2026ema}.
On the lattice, the normal ordering is a finite, $ga$-dependent multiplicative renormalization,
\begin{equation}
:\!e^{\pm 2i\sqrt{\pi}\phi_j}\!:_{\infty}
= \mathcal{N}(ga)\, e^{\pm 2i\sqrt{\pi}\phi_j},
\qquad
\mathcal{N}(ga) = \exp\!\left[
\frac{K\!\left(\bigl(1+\tfrac{(ga)^2}{4\pi}\bigr)^{-1}\right)}
     {\sqrt{1+\tfrac{(ga)^2}{4\pi}}}
\right],
\label{eq:normal-ordering}
\end{equation}
with $K$ the complete elliptic integral of the first kind, in the
parameter convention defined in Appendix~\ref{sec:freecovariances}.
Exponentials $e^{i\lambda\phi}$, with real $\lambda$, are called vertex operators.
With $z=ga$, the combination $z\mathcal{N}(z)$ has an expansion in even powers of $z$ whose coefficients are polynomials in $L(z)=\ln(8\sqrt\pi/z)$,
\begin{equation}
z\mathcal N(z)=8\sqrt{\pi}\left[
1+\frac{z^2}{16\pi}(1-L)
+\frac{z^4}{2048\pi^2}(4L^2+10L-17)
+O\!\left(z^6L^3\right)
\right]. \label{eq:normal-ordering-expansion}
\end{equation}
The corresponding treatment of normal ordering on a Euclidean spacetime lattice is discussed in~\cite{Bender:1984qg,Ohata:2023sqc}.
The factor in Eq.~\eqref{eq:normal-ordering} uses the equal-time covariance of the continuous-time Hamiltonian~\eqref{eq:Hlat}, evaluated in Appendix~\ref{sec:freecovariances}.
The free
lattice theory is exactly solvable, with dispersion
\begin{equation}
\Omega(k)=\frac{1}{a}\sqrt{4\sin^2\!\left(\frac{ka}{2}\right)+\frac{(ga)^2}{\pi}}\,.
\label{eq:dispersion}
\end{equation}
Here $k$ is the spatial momentum and $\Omega(k)$ is the normal-mode frequency. 
Its $k\to0$ limit is the meson mass $g/\sqrt\pi$ exactly at every lattice spacing. 
We set $\theta=\pi$
throughout.
Numerical results below use units $g=1$: quoted masses and lattice
spacings are $m/g$ and $ga$, respectively.

\subsection{Truncation of the local Hilbert space}

A uniform MPS is a state built by repeating the same MPS tensor along an infinite chain.
For each local basis state $|n\rangle$, the tensor contains a $D\times D$ matrix $A^n$.
We write the uniform MPS as
\[
 |\Psi(A)\rangle
 = \sum_{\{n_j\}}
   \bigl(\cdots A^{n_{-1}}A^{n_0}A^{n_1}\cdots\bigr)
   |\cdots n_{-1}n_0n_1\cdots\rangle .
\]
The sum runs over the local basis labels at every site, and the parentheses denote contraction of all bond indices.
The same matrices $A^n$ appear at every site.
This infinite-chain notation is understood through a thermodynamic limit: on a periodic chain of $N$ sites, the coefficient of $|n_1\cdots n_N\rangle$ is $\operatorname{tr}(A^{n_1}\cdots A^{n_N})$, and local expectation values are obtained by taking $N\to\infty$ after normalizing the finite-chain state.
Cutting a bond divides the chain into two halves. The Schmidt decomposition expresses the state as a weighted sum of products of orthonormal states on these halves.
The bond dimension $D$ bounds the number of terms in this sum and therefore limits the entanglement that the MPS can represent~\cite{Perez-Garcia:2006nqo}.
Given the lattice Hamiltonian, the VUMPS algorithm iteratively optimizes the matrices $A^n$ at fixed bond dimension $D$ to minimize the energy per site~\cite{ZaunerStauber:2018vumps}.
The resulting uniform MPS is a variational approximation to the ground state.

For the bosonic chain, we must also choose a truncation of the infinite-dimensional Hilbert space at each site.
The oscillator eigenstates of the on-site term $p_j^2/(2a)+a\mu^2\phi_j^2/2$, with $\mu=g/\sqrt{\pi}$, are a possible choice. 
They ignore the coupling between sites and require many levels as $a\to0$.
We instead use the eigenstates of the one-site reduced density matrix of the free ($m=0$) chain. 
These states account for the coupling through the gradient term $(\phi_{j+1}-\phi_j)^2$.
Writing $\varepsilon := (ga)^2/4\pi$ and
$\tilde m := 1/(1+\varepsilon)$, the free-chain variances are
\begin{equation}
\sigma_\phi^2 = \langle\phi_j^2\rangle_0
= \frac{K(\tilde m)}{2\pi\sqrt{1+\varepsilon}}
= \frac{\ln\mathcal{N}(ga)}{2\pi},
\qquad
\sigma_p^2 = \langle p_j^2\rangle_0
= \frac{2}{\pi}\sqrt{1+\varepsilon}\;E(\tilde m).
\label{eq:variances}
\end{equation}
Here $E$ is the complete elliptic integral of the second kind, and
the subscript $0$ denotes the free infinite-chain vacuum.
The free ground-state energy per site, which is its zero-point energy, is $E_{\rm per\,site}=\sigma_p^2/a$.
Appendix~\ref{sec:freecovariances} derives these results by evaluating momentum integrals.

Tracing out all sites except $j$ gives a reduced density matrix $\rho_1$. 
Although the full vacuum is pure, $\rho_1$ is mixed because site $j$ is entangled with the rest of the chain. 
The standard calculation for coupled harmonic oscillators~\cite{Srednicki:1993im,Peschel:2009rdm,Audenaert:2002xfl} shows that its eigenfunctions are harmonic-oscillator wave functions with a frequency fixed by the two variances~\eqref{eq:variances}.
In our conventions, the corresponding basis is defined by
\begin{equation}
\phi_j = \ell_0\,(b_j+b_j^\dagger),
\qquad
\ell_0^2 = \frac{1}{2a\omega_0},
\qquad
\omega_0 := \frac{1}{a}\sqrt{\frac{\sigma_p^2}{\sigma_\phi^2}}
\simeq \frac{2}{a}\Bigl[\ln\frac{8\sqrt{\pi}}{ga}\Bigr]^{-1/2}.
\label{eq:omega0}
\end{equation}
The operators $b_j$ and $b_j^\dagger$ annihilate and create a quantum of this local oscillator, with $[b_j,b_l^\dagger]=\delta_{jl}$ and $b_j^\dagger b_j|n\rangle=n|n\rangle$. The parameter $\ell_0$ fixes the width of its wave functions.
The frequency scales as $a^{-1}$ times a slowly varying logarithmic factor; over $0.1\le ga\le0.8$, $a\omega_0$ ranges from approximately $0.90$ to $1.23$.
It sets the width of the local basis functions and need not equal the physical mass $\mu$. 
The eigenvalues of the reduced density matrix are geometrically distributed:
\begin{equation}
\rho_1 = (1-q)\sum_{n=0}^{\infty}q^n\ket n\bra n,
\qquad
q = \frac{\bar n}{1+\bar n},
\qquad
\bar n := \sigma_\phi\sigma_p-\tfrac12
=\frac{\sqrt{K(\tilde m)E(\tilde m)}}{\pi}-\tfrac12.
\label{eq:thermal}
\end{equation}
Here $\bar n=\operatorname{Tr}(\rho_1 b_j^\dagger b_j)$ is the mean occupation number.
These eigenvalues have the form of a thermal oscillator distribution, but no physical temperature is introduced. 
Keeping the $n_{\max}+1$ largest eigenvalues retains the greatest possible probability for a local subspace of that dimension. 
The discarded probability is $q^{n_{\max}+1}$.
Because $\bar n$ grows only like $\sqrt{\ln(1/a)}$, a discarded probability below $10^{-10}$ requires $n_{\max}=8$--$13$ over the full range $ga=0.8$--$0.1$.
For $n_{\max}=30$, the corresponding free-theory tail is at most $2\times10^{-24}$, whereas the interacting cutoff checks in Sec.~\ref{sec:budget} find occupation probabilities as large as $2\times10^{-9}$ beyond level 30; the numerical cutoff test is therefore essential.
This construction is an analytic counterpart, exact at $m=0$, of the numerically optimized local bases in Refs.~\cite{Zhang:1998lbo,Brockt:2015lbo}.
For $m\ne0$, we keep $\omega_0(ga)$ at its free-theory value and test the local truncation numerically.
We increase $n_{\max}$ from 30 to 34 at four lattice spacings with $D=192$; Sec.~\ref{sec:budget} reports the resulting changes.

\subsection{Matrix elements and truncation checks}

We retain $\ket n$ with $0\le n\le n_{\max}=30$ at every lattice spacing. 
Beyond convergence of the state, the cutoff must resolve the vertex operator in the mass interaction.
To see the additional requirement, write $\phi=\ell(b+b^\dagger)$ with a positive scale $\ell$, and let $\ket0$ denote the vacuum of this local oscillator.
The exact matrix elements give the Poisson distribution $|\langle n|e^{i\lambda\phi}|0\rangle|^2=e^{-s}s^n/n!$, where $s=\lambda^2\ell^2$. 
Applying the vertex to $\ket0$ therefore populates levels around $n=s$, with a spread $\sqrt{s}$.
Resolving this distribution requires levels up to several standard deviations above $s$. 
This estimates the required cutoff.

For the mass operator, $\lambda=\beta:=2\sqrt{\pi}$, the chosen basis gives $s_\beta=4\pi\ell_0^2\approx\pi\sqrt{\ln(8\sqrt{\pi}/ga)}$.
It grows only as the square root of a logarithm, from $\approx5.1$ at $ga=0.8$ to $\approx7.0$ at $ga=0.1$.
In the on-site basis with frequency $\mu$, it instead grows as $2\pi^{3/2}/ga$, from $\approx14$ to $\approx111$ over the same range.
The eigenbasis of the free-chain one-site reduced density matrix therefore needs far fewer states than the on-site basis at fine spacings.

We construct the interaction from the exact harmonic-oscillator matrix elements~\cite{Cahill:1969iq},
\begin{equation}
\braket{m'|e^{i\lambda\phi}|n}
=e^{-s/2}(i\lambda\ell)^{m'-n}
\sqrt{\frac{n!}{m'!}}\;L_n^{(m'-n)}(s),
\qquad m'\ge n.
\label{eq:displacement}
\end{equation}
Here $m'$ and $n$ are nonnegative oscillator quantum numbers, and $L_n^{(\alpha)}(s)$ is the associated Laguerre polynomial with parameter $\alpha=m'-n$.
For $m'<n$, the matrix element follows from the Hermitian-conjugation relation $\langle m'|e^{i\lambda\phi}|n\rangle=\langle n|e^{-i\lambda\phi}|m'\rangle^*$.
We retain their exact finite block rather than exponentiating the truncated field matrix. 
The latter omits intermediate oscillator levels and changes the operator.
The cosine connects levels of the same parity, and the sine connects levels of opposite parity.

\subsection{Charge conjugation and the transfer matrix}

At $\theta=\pi$, the Hamiltonian is invariant under $\phi_j\to-\phi_j$ and $p_j\to-p_j$.
Charge conjugation acts on each site as oscillator parity, $C=(-1)^{b_j^\dagger b_j}$. 
We assign even and odd parity to the tensor indices and allow only components that conserve this $\mathbb Z_2$ charge~\cite{Singh:2009cd}. 
This keeps the variational state invariant under $C$; we call the resulting family of solutions the symmetric branch.

Correlation functions are obtained from the MPS transfer matrix, formed by contracting the physical index of a tensor with that of its complex conjugate. 
Normalize its largest eigenvalue to one. 
If $\lambda_r$ is another eigenvalue, a separation of $L$ sites gives a contribution proportional to $\lambda_r^L$, so $\epsilon_r=-\ln|\lambda_r|$ is an inverse correlation length in lattice units~\cite{Zauner:2014iea}.
Charge conjugation separates these eigenvalues into even and odd sectors.

In a canonical MPS representation, the bond bases correspond to orthonormal Schmidt states on the two sides of a cut.
To see how their parities determine the transfer-matrix sectors, let $P$ be the diagonal matrix whose entries $\pm1$ give the parity of each bond index.\footnote{For an injective $C$-invariant MPS in canonical form, the symmetry relation is $(-1)^nA^n=\eta V^\dagger A^nV$, with $V$ unitary and $\eta=\pm1$~\cite{Perez-Garcia:2008hhq}.
Injectivity means that, for some finite length $L$, the products $A^{n_1}\cdots A^{n_L}$ span all $D\times D$ matrices.
Since $C^2=1$, injectivity implies $V^2=e^{i\alpha}I$, where $I$ is the identity matrix and $\alpha$ is real; replacing $V$ by $e^{-i\alpha/2}V$ and changing the bond basis gives $V=P=\operatorname{diag}(\pm1)$.
This leaves $\eta$ unchanged.
On a periodic chain of $N$ sites, the total $C$ eigenvalue is $\eta^N$.
Our charge-conserving tensors implement $\eta=+1$.
This choice matches the finite-volume vacuum: before truncating the local Hilbert space, the Hamiltonian has a real, confining potential invariant under simultaneous reversal of all $\phi_j$.
Its ground-state wavefunction is unique and can be chosen strictly positive, so it is $C$-even for every $N$.
The sign $\eta$ cancels from the transfer matrix and hence does not affect its even/odd block decomposition.}
Keeping only the
components that conserve the $\mathbb Z_2$ charge means
\begin{equation}
P A^n P = (-1)^n A^n .
\label{eq:gradedtensor}
\end{equation}
Applying $C$ to every site replaces each $A^n$ by $PA^nP$. 
The two factors of $P$ that meet on each bond multiply to $P^2=1$, so the state is invariant. 
The transfer matrix $E=\sum_n A^n\otimes\bar A^n$ acts on pairs of bond indices, and Eq.~\eqref{eq:gradedtensor} gives $(P\otimes P)E(P\otimes P)=E$.
The bar denotes entrywise complex conjugation.
Hence $E$ is block diagonal. 
Its even block connects bond pairs of equal parity and contains the dominant eigenvector; its odd block connects pairs of opposite parity. 
A local operator $O$ enters a correlation function through $E_O=\sum_{n,n'}O_{nn'}A^{n'}\otimes\bar A^{n}$. 
If \(O\) is odd under \(C\), its matrix element \(O_{nn'}=\langle n|O|n'\rangle\) vanishes unless one of \(n,n'\) is even and the other is odd. Consequently, \(E_O\) maps the even sector of the transfer matrix into the odd sector, and the subsequent powers of \(E\) act within the odd sector.
For operators at sites $j$ and $l$, the connected correlator is $\langle O_jO_l\rangle_c=\langle O_jO_l\rangle-\langle O_j\rangle\langle O_l\rangle$.
The connected correlator of a $C$-odd operator therefore decays with the odd-sector eigenvalues alone, and that of a $C$-even operator with the even sector. 
A $C$-odd operator can detect the broken symmetry if its expectation value is nonzero in the broken phase.
A conventional choice is $\sin(2\sqrt{\pi}\phi_j)$, the bosonized pseudoscalar density $\bar\psi i\gamma_5\psi$, where $\gamma_5$ is the chirality matrix.
At $\theta=\pi$, $\partial H_{\rm lat}/\partial\theta$ is proportional to its sum over sites; its expectation value therefore measures the slope of the vacuum energy in $\theta$, which jumps across the first-order line for $m/g>(m/g)_c$~\cite{Coleman:1976uz}.
On the symmetric branch its expectation value vanishes.
Its correlation length equals the largest odd-sector correlation length when the leading odd-sector mode contributes with a nonzero coefficient to its correlation function.
At each bond dimension $D$, let $\lambda^{\rm odd}_1$ and $\lambda^{\rm odd}_2$ be the eigenvalues of the odd block with the largest and second-largest moduli, respectively, so that $|\lambda^{\rm odd}_1|\geq|\lambda^{\rm odd}_2|$.
We define
\[
 \epsilon_1=-\ln|\lambda^{\rm odd}_1|,
 \qquad
 \epsilon_2=-\ln|\lambda^{\rm odd}_2|,
 \qquad \epsilon_1\leq\epsilon_2.
\]
We use an iterative eigensolver to compute the four eigenvalues of largest modulus in the odd sector. We then evaluate $-\ln|\lambda|$ for each, sort these values in ascending order and retain the first two.
The transfer matrix is constructed from the canonical MPS tensor, which fixes its dominant even-sector eigenvalue to one.
Our unit cell contains one site, so these logarithms are inverse correlation lengths in lattice units.
We use $\xi_{\rm odd}=1/\epsilon_1$ and $\delta=\epsilon_2-\epsilon_1$; the physical correlation length is $a\xi_{\rm odd}$.
The subscript $\infty$ on $\epsilon_{1,\infty}$ or $\xi_{{\rm odd},\infty}$ denotes the limit $D\to\infty$.
For an order parameter whose correlations at large distances are governed by the leading odd-sector mode, the transition is signaled by $\epsilon_{1,\infty}\to0$ after extrapolation to infinite bond dimension.
 
Enforcing $C$ exactly also prevents a finite-$D$ optimization near the transition from selecting a slightly $C$-broken state, for which the eigenvalues would carry no parity label and neither $\xi_{\rm odd}$ nor $\delta$ would be defined.

\section{Critical-point estimators}
\label{sec:fes}

\subsection{The effect of finite bond dimension}
\label{sec:seq}

Finite bond dimension limits the correlation length that the
variational state can represent. It can therefore shift the apparent
critical point even though the chain is infinite. We first describe
this effect and then explain how we remove it.

We increase the bond dimension in steps, initializing each calculation with a preceding converged state.
The largest dimensions range from $192$ to $320$, as listed in Table~\ref{tab:perspacing}.
We write $m_*(a)$ for the critical mass at lattice spacing $a$, after extrapolating to infinite bond dimension.
At fixed lattice spacing \(a\), Ising critical scaling predicts that the correlation length of the exact ground state diverges as \(\xi_{\mathrm{odd},\infty}\propto[(m_*(a)-m)/g]^{-\nu}\), with \(\nu=1\), as \(m\) approaches \(m_*(a)\) from the symmetric side.
Its inverse therefore vanishes linearly to leading order in \(m_*(a)-m\).
At finite $D$, the variational state has an additional finite correlation length, which must be extrapolated out.

One possible extrapolation uses the finite-entanglement law $\xi_D\propto D^\kappa$~\cite{Tagliacozzo:2008fes,Pollmann:2009lnv}.
Here $\xi_D$ is the correlation length of the finite-$D$ variational state at criticality, and $\kappa$ is its asymptotic growth exponent.
Using this law to extrapolate a critical-mass estimate $m_*(a,D)$ requires bond dimensions large enough to reach the asymptotic regime~\cite{Vanhecke:2019pez}.
Our data give approximately $\xi_D\propto D^{1.7}$, below the Ising exponent $\kappa=2.03$.
We therefore use the transfer-matrix spectrum itself to measure the finite-$D$ effect.

\subsection{Extrapolation in the transfer-matrix spectrum}
\label{sec:delta}

A finite bond dimension introduces a length scale into the state, as a finite system size does in finite-size scaling.
We measure it by the spacing of the two lowest levels of the transfer matrix in the $C$-odd sector, $\delta=\epsilon_2-\epsilon_1$, which vanishes as $D\to\infty$~\cite{Rams:2018lnq,Vanhecke:2019pez}. 
Near the critical point, the departure from criticality in the mass direction is described by a signed parameter $t$, called a scaling field. At fixed lattice spacing, its leading behavior is $t=A(a)[m_*(a)-m]+O([m-m_*]^2)$, with $A(a)>0$. Thus $t$ vanishes at the critical mass, is positive in the symmetric phase, and is negative in the broken-symmetry phase. We choose its normalization so that it agrees with the leading critical behavior of the inverse correlation length $\epsilon_{1,\infty}$ on the symmetric side. Unlike $t$, the physical inverse correlation length is nonnegative.
Under coarse-graining by a length factor $b$, the scaling field transforms as $t\mapsto b^{1/\nu}t$ to leading order. For the Ising exponent $\nu=1$, $t$ and $\delta$ therefore both scale as inverse lengths, giving the finite-entanglement scaling form
\begin{equation}
\epsilon_1/\delta = F(x),\qquad x=t/\delta
\label{eq:scalinglaw}
\end{equation}
for some function~$F(x)$~\cite{Vanhecke:2019pez}.
At the critical mass, $t=0$, so Eq.~\eqref{eq:scalinglaw} gives
$\epsilon_1=F(0)\delta$
within the leading scaling approximation.
We locate the mass where this proportionality holds over the fitted bond dimensions and test the effects of corrections below.

Schneider et al.~\cite{Schneider:2024ipg} interpret the limiting
transfer-matrix ratios at criticality using an effective lattice model.
For symmetry-preserving MPS tensors, this is an open Ising chain whose
transverse field differs from its critical value by an amount
proportional to the inverse chain length.
The effective chain length grows with the MPS correlation length as $D$ increases.
The deviation of the transverse field from criticality decreases as the inverse length, but its product with the length approaches a nonzero constant.
Thus the perturbation still affects the lowest energy levels in the limit of a long chain. This limiting regime is their self-congruent point.
The lowest energy-gap ratios of this model approximately reproduce
the MPS transfer-matrix ratios, providing a quantitative estimate
of the critical slope $F(0)$.
The dependence of $F(x)$ away from criticality is instead described
by the scaling hypothesis in Eq.~\eqref{eq:scalinglaw}.

To choose the mass range at each lattice spacing, we use the largest bond dimension for which at least three input masses were simulated. Suppose the $N$ masses at this bond dimension run from $m_{\min}$ to $m_{\max}$, with average separation $s=(m_{\max}-m_{\min})/(N-1)$. We include masses from $m_{\min}-s$ to $m_{\max}+s$: one additional average mass step at either end. We then fit all computed points with $D\ge128$ whose masses fall in this interval to
\begin{equation}
\epsilon_1(m,D)=p_1\,[m-m_*(a)]+c_1\,\delta(m,D).
\label{eq:jointfit}
\end{equation}
Here $p_1$ is the slope with respect to $m$ at fixed $\delta$, and $c_1$ is the slope with respect to $\delta$ at fixed $m$; both are shared by all fitted points at that spacing.
The fit determines the critical mass $m_*(a)$ directly, as the mass at which the $\delta=0$ intercept vanishes. Starting from this estimate, we further restrict the selected points to $|m-m_*|\le6\times10^{-4}$ and refit, repeating this step twice. The primary fits use 257 of the 500 archived points.
 
\begin{figure}[!htbp]
\centering
\includegraphics[width=\textwidth]{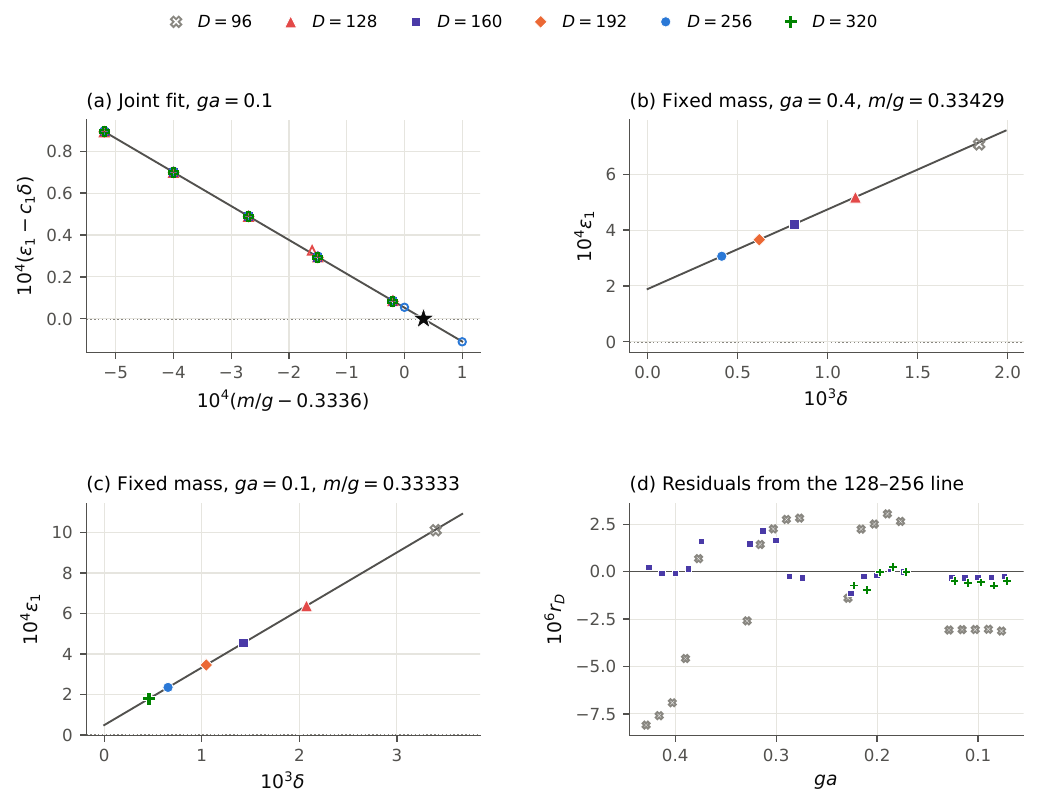}
\caption{Removing the finite-bond-dimension error using the
transfer-matrix spacing $\delta$.
(a) The joint fit, Eq.~\eqref{eq:jointfit}, at $ga=0.1$: subtracting $c_1\delta$ leaves the linear mass dependence, whose zero, marked by the star, determines $m_*(a)$. 
After the subtraction, the points at different bond dimensions lie on top of one another; the markers are open so that all of them remain visible.
(b,c) $\epsilon_1$ against $\delta$ at a fixed mass, $ga=0.4$ and
$0.1$: the lines are fitted to $D\ge128$ (up to $D=320$ at $ga=0.1$)
and show the linearity assumed in Eq.~\eqref{eq:jointfit}. The
critical mass is not extracted from these panels.
(d) Residuals $r_D=\epsilon_1(D)-\ell_{128,256}(\delta_D)$ at
$D=96$, $160$ and, at $ga=0.2$ and $0.1$, $320$, where
$\ell_{128,256}$ is the line through $D=128$ and $256$ at the same
mass. Each group contains the five fixed masses in increasing order
from left to right; small horizontal offsets separate the bond
dimensions.
The $D=96$ data (gray) test linearity and are excluded from the fits.}
\label{fig:finiteD}
\end{figure}

We call Eq.~\eqref{eq:jointfit} the joint fit. 
It uses every point retained by this selection, whether or not the same mass was computed at every bond dimension.
At a fixed mass, the fit is a straight line in $\delta$ (Fig.~\ref{fig:finiteD}(b,c)). 
Within the first-order expansion discussed below, its $\delta=0$ intercept is proportional to the signed scaling field $t$ and therefore vanishes at the critical mass.
The joint fit imposes linear vanishing in $m$, as expected for the Ising exponent $\nu=1$ and found at every lattice spacing studied in Ref.~\cite{Byrnes:2002nv}; the quadratic-intercept variation in Table~\ref{tab:continuum-checks} tests the leading departure from this assumption.

Reference~\cite{Fujii:2024giv} uses a two-stage method: it extrapolates $\epsilon_1$ in $\delta$ at each fixed mass, then finds the mass at which the fitted intercept vanishes.\footnote{Reference~\cite{Fujii:2024giv} motivates the linear extrapolation on the symmetric side using the regime $x=\epsilon_{1,\infty}/\delta\gg1$.
There the transfer-matrix levels resolve the continuum above $\epsilon_{1,\infty}$, and a linear extrapolation in $\delta$ approximately recovers $\epsilon_{1,\infty}$ itself~\cite{Zauner:2014iea,Rams:2018lnq}.
Their data, however, have $\epsilon_1/\delta\le1.7$, corresponding to $x\lesssim1.4$ under the empirical Ising-chain interpolation in Eq.~\eqref{eq:tfimF}.
Near criticality, the signed small-$x$ expansion instead gives $\epsilon_1\approx F_1t+F_0\delta$, with $F_0=F(0)$ and $F_1=F'(0)$.
The fitted intercept then represents $F_1t$. This factor changes its magnitude but leaves its zero, and hence the extracted critical mass, unchanged within the first-order approximation.
The Ising-chain test in Appendix~\ref{sec:tfim} supports this interpretation and gives $F_1\simeq1.05$.}
For our data, at a fixed mass, the $D=160$ and $D=320$ points lie within $2.2\times10^{-6}$ and $10^{-6}$ of the line through the $D=128$ and $D=256$ points, whereas the $D=96$ points deviate by up to $9.5\times10^{-6}$; we therefore use $D\ge128$ in the fits.
At the five spacings of Table~\ref{tab:seqcmp}, the two-stage estimate agrees with the joint fit within $4\times10^{-6}$. 
It allows a separate slope at each mass, and the difference largely reflects the mass dependence of the slope discussed next.

The joint fit follows by expanding the scaling function near criticality:
$F(x)=F_0+F_1x+F_2x^2+F_3x^3+\cdots$, with $F_k=F^{(k)}(0)/k!$.
Keeping the first two terms in Eq.~\eqref{eq:scalinglaw} gives $\epsilon_1\simeq F_0\delta+F_1t$.
Keeping only the linear mass dependence of $t$ then gives Eq.~\eqref{eq:jointfit}, with $p_1=-F_1A(a)$ and $c_1=F_0$ within this approximation.
The fitted $\delta=0$ intercept is therefore $F_1t$ to first order in $m-m_*(a)$ and can change sign across the transition. It extrapolates the small-$|x|$ expansion, rather than taking the physical limit $D\to\infty$ at fixed mass. On the symmetric side it differs from the leading inverse correlation length by the factor $F_1$; its zero nevertheless gives the critical mass within this approximation.
Using $F_1\simeq1.05$ from the Ising-chain interpolation in Appendix~\ref{sec:tfim}, we estimate $x\simeq p_1[m-m_*(a)]/(F_1\delta)$ for the selected points. They span approximately $-0.17\le x\le1.22$. On this estimate, their negative-$x$ range lies within the interpolation domain $x\ge-0.21$ used in that appendix.
A free-fermion version of the self-congruent point gives $\epsilon_2/\epsilon_1=4.60$ for the Ising class~\cite{Schneider:2024ipg}, i.e.\ $F_0=0.278$, against $c_1=0.279$--$0.287$ in our fits and $0.279$ for the transverse-field Ising chain (Appendix~\ref{sec:tfim}).
Since that picture reproduces only the lowest levels and the fitted slopes vary by a few percent between spacings, we use the value as a check and not as an input.
At the critical mass, $x=0$ and the terms of order $x^2$ and higher vanish within the leading scaling form.
Away from that mass they change the slope against $\delta$ and can affect the fitted zero.
Departures from Eq.~\eqref{eq:scalinglaw} at finite bond dimension can remain even at the critical mass.
 
We measured the effect of the higher-order terms in $F(x)$ in two ways. 
First, we added the next two orders, which with $u=m-m_*(a)$ read 
\begin{equation}
\epsilon_1=p_1u+c_1\,\delta+\phi_2\,\frac{p_1^2u^2}{\delta}
+\phi_3\,\frac{p_1^3u^3}{\delta^2},\qquad \phi_k=\frac{F_k}{F_1^k}.
\label{eq:extfit}
\end{equation}
We fit $\phi_2$ and $\phi_3$ as common coefficients across lattice spacings, as suggested by the limiting Ising scaling function.
This is an approximation at finite $a$ and $D$. In this fit, residuals at each spacing are divided by fixed reference rms values, which are supplied with the analysis code.
 
Second, we fitted synthetic data generated on our masses and bond dimensions with the scaling function of the transverse-field Ising chain (Appendix~\ref{sec:tfim}).
Both tests indicate upward shifts of $m_*(a)$ of at most $1.9\times10^{-6}$ at the coarsest spacings and $0.9\times10^{-6}$ at $ga\le0.25$.
The fitted coefficients and the resulting changes in the continuum estimate are reported in Sec.~\ref{sec:budget}.

\FloatBarrier
\section{Continuum analysis and results}
\label{sec:results}

\subsection{Per-spacing results}
\label{sec:perspacing}

Table~\ref{tab:perspacing} lists the critical masses obtained at
eleven lattice spacings, $0.1\le ga\le0.8$, using the joint fit
described in Sec.~\ref{sec:delta}.
These masses and their estimated numerical uncertainties are
the inputs to the continuum extrapolation.

\begin{table}[htb]
\centering
\begin{tabular}{lllc}
\toprule
$ga$ & $m_*(a)$ & $\sigma_a\,[10^{-6}]$ & bond dimensions $D$ \\
\midrule
0.8   & 0.337\,247 & 1.2 & 128--192 \\
0.7 & 0.336\,452 & 1.1 & 128--256 \\
0.6   & 0.335\,738 & 1.4 & 128--192 \\
0.5   & 0.335\,114 & 1.1 & 128--256 \\
0.4   & 0.334\,585 & 1.5 & 128--256 \\
0.3   & 0.334\,161 & 2.2 & 128--256 \\
0.25  & 0.333\,987 & 2.2 & 128--256 \\
0.2   & 0.333\,839 & 1.6 & 128--320 \\
0.15  & 0.333\,724 & 2.1 & 128--320 \\
0.125 & 0.333\,672 & 1.4 & 128--320 \\
0.1   & 0.333\,633 & 1.8 & 128--320 \\
\bottomrule
\end{tabular}
\caption{Critical masses $m_*(a)$ from the joint fit with $D\ge128$ (Sec.~\ref{sec:delta}). 
The uncertainties~$\sigma_a$ are defined in Eq.~\eqref{eq:input-error} from the joint-fit residuals and the changes under variations of the bond dimensions and mass dependence.
At $ga=0.4$, $0.3$, $0.2$, $0.15$,
$0.125$ and $0.1$, five fixed masses were computed at every bond
dimension, including $D=160$; at $ga=0.15$ and $0.125$, one of the five
$D=320$ calculations did not reach the solver tolerance
$3\times10^{-10}$ and is not used.
Converged runs at $m=0.33361$ and $0.33368$, respectively, replace these points in the joint fits and the prediction check of Sec.~\ref{sec:budget}, giving five usable $D=320$ points at each spacing.}
\label{tab:perspacing}
\end{table}

For each lattice spacing $a$, we estimate the numerical uncertainty
$\sigma_a$ in the critical mass $m_*(a)$ by combining three
contributions in quadrature:
\begin{equation}
 \sigma_a^2=\sigma_{{\rm LS},a}^2+\Delta_{D,a}^2+\Delta_{m,a}^2.
 \label{eq:input-error}
\end{equation}
The first contribution is obtained from the residuals of the unweighted joint fit, Eq.~\eqref{eq:jointfit}.
The subscript $\mathrm{LS}$ stands for least squares.
Let $i=1,\ldots,N_a$ label the points $(m_i,D_i)$ used in that fit, and write
$\epsilon_{1,i}=\epsilon_1(m_i,D_i)$ and $\delta_i=\delta(m_i,D_i)$.
The residuals and the derivatives of the fitted function with respect to the three parameters $(m_*,p_1,c_1)$ are
\begin{equation}
 \begin{aligned}
 r_i&=\epsilon_{1,i}-p_1[m_i-m_*(a)]-c_1\delta_i,\\
 (J_{i1},J_{i2},J_{i3})&=(-p_1,\;m_i-m_*(a),\;\delta_i),
 \end{aligned}
 \label{eq:jointfit-residuals}
\end{equation}
evaluated at the fitted parameter values.
The matrix $J$ collects these derivatives; a superscript $\mathsf T$ denotes matrix transposition.
Using the residual variance $\sum_i r_i^2/(N_a-3)$ and propagating it to the fitted mass gives
\begin{equation}
 \sigma_{{\rm LS},a}^2
 =\frac{\sum_{i=1}^{N_a}r_i^2}{N_a-3}
   \bigl[(J^{\mathsf T}J)^{-1}\bigr]_{11}.
 \label{eq:jointfit-ls-error}
\end{equation}
The matrix element selects the variance of the first parameter, $m_*$, in the linear approximation to the parameter dependence.

For the bond-dimension comparison, let $m_*^{(-D_{\max})}(a)$ be the result of repeating the joint fit after removing all points at the largest available bond dimension, and let $m_*^{(96)}(a)$ be the result with $D\ge96$ instead of $D\ge128$.
We define
\begin{equation}
 \Delta_{D,a}=
 \begin{cases}
  \bigl|m_*^{(96)}(a)-m_*(a)\bigr|,&ga=0.8\text{ or }0.6,\\
  \bigl|m_*^{(-D_{\max})}(a)-m_*(a)\bigr|,&\text{otherwise}.
 \end{cases}
 \label{eq:bond-dimension-shift}
\end{equation}
At $ga=0.8$ and $0.6$, only $D=128$ and $192$ enter the central fit, so omitting $D=192$ would leave only one bond dimension.
These two spacings provide a weaker test of the bond-dimension dependence.
We therefore examine the effect of omitting the coarsest spacings in the continuum extrapolation.

For the mass-dependence comparison, we replace Eq.~\eqref{eq:jointfit} by
\begin{equation}
 \epsilon_1(m,D)=p_1^{\rm quad}[m-m_*^{\rm quad}(a)]
 +p_2^{\rm quad}[m-m_*^{\rm quad}(a)]^2
 +c_1^{\rm quad}\delta(m,D),
 \label{eq:jointfit-quadratic}
\end{equation}
with four fitted parameters.
If the quadratic polynomial at $\delta=0$ has two real zeros, $m_*^{\rm quad}(a)$ denotes the one nearest the mean mass of the points used in the fit.
The third contribution is
\begin{equation}
 \Delta_{m,a}=\bigl|m_*^{\rm quad}(a)-m_*(a)\bigr|.
 \label{eq:mass-dependence-shift}
\end{equation}
For each modified fit, we select data satisfying \(|m-m_*(a)|\le6\times10^{-4}\), updating the estimated critical mass and repeating the selection and fit twice.
The Zenodo release~\cite{Okuda:2026bosonizeddata} gives the unrounded masses and all three components of Eq.~\eqref{eq:input-error}.

\subsection{Fitting functions and the Akaike-weighted average}
\label{sec:contmodel}

The continuum analysis has three steps. We fit several functions to the critical masses in Table~\ref{tab:perspacing}, retain fits that meet a common residual criterion, and average their continuum intercepts.
The uncertainty of this average accounts for both the input-mass uncertainties and the spread among the retained intercepts. 
We define each step below.

The free lattice dispersion, Eq.~\eqref{eq:dispersion}, has a leading correction of order $a^2$.
The expansion of $z\mathcal{N}(z)$, Eq.~\eqref{eq:normal-ordering-expansion}, contains logarithms, but contributions from this factor can cancel against other terms when a physical quantity is matched.
Logarithms can arise in the coefficients of local corrections to the continuum action when its correlation functions are required to reproduce those of the lattice theory at distances much larger than $a$, order by order in $a$; their presence and coefficients must be determined by the complete matching calculation.
This procedure is called matching and is described in Sec.~\ref{sec:localmatching}.

For the primary estimate we use $z=ga$ and fit
\begin{equation}
 m_*(a)=c_0+c_2z^2+c_{2L}z^2\ln z+R_q(z),\qquad c_0=(m/g)_c.
 \label{eq:family}
\end{equation}
We specify $R_q$ by six nested polynomials in $z^2$,
\begin{equation}
 R_q(z)=\sum_{j=2}^{q+1}c_{2j}z^{2j},\qquad q=0,1,\ldots,5,
 \label{eq:family-additional}
\end{equation}
so that $R_0=0$ and the six fitting functions end at $z^2,z^4,\ldots,z^{12}$.
The first function has the three coefficients $c_0$, $c_2$ and $c_{2L}$.
Each further function adds one of $c_4$, $c_6$, $c_8$, $c_{10}$ and $c_{12}$, in this order.
All displayed coefficients, including $c_{2L}$, are independent fit parameters, so the function with index $q$ has $q+3$ of them.
Replacing $\ln z$ by $\ln(z/z_0)$ with a constant $z_0>0$ is equivalent to replacing $c_2$ by $c_2-c_{2L}\ln z_0$, which is a change of variables among the free parameters.
The family therefore does not depend on the scale in the logarithm.

The power counting in Sec.~\ref{sec:physicalcounting} allows even powers of $z$, but does not fix the highest power of $\ln z$ at each order.
Our primary family makes two additional choices: it keeps only $c_2+c_{2L}\ln z$ at order $z^2$, and omits logarithms at order $z^4$ and higher.
Section~\ref{sec:logcheck} explains the assumption about coefficient sizes behind this omission.
It also tests functions with logarithms at order $z^4$, a term $z^2(\ln z)^2$, or a different highest power of $z$.
Each fitting function is applied to all eleven entries in Table~\ref{tab:perspacing} and to four subsets, defined by $z\le0.6$, $z\le0.5$, $z\le0.4$ and $z\le0.3$.
For each index $q$ and set of included spacings $\mathcal S$, we determine the coefficients $\boldsymbol c=(c_0,c_2,c_{2L},c_4,\ldots,c_{2q+2})$ by minimizing
\begin{equation}
 \chi^2(\boldsymbol c)
 =\sum_{a\in\mathcal S}
 \frac{\left[m_*(a)-c_0-c_2(ga)^2-c_{2L}(ga)^2\ln(ga)-R_q(ga)\right]^2}
 {\sigma_a^2},
 \label{eq:continuum-weighted-residuals}
\end{equation}
where $R_q$ is defined in Eq.~\eqref{eq:family-additional}.
The per-spacing joint fits described in Sec.~\ref{sec:delta} determine the critical-mass estimates $m_*(a)$ listed in Table~\ref{tab:perspacing}.
These estimates and their numerical uncertainties $\sigma_a$, defined in Eq.~\eqref{eq:input-error}, are then used as fixed data in the continuum fit.
Each squared residual has weight $1/\sigma_a^2$, so a mass estimate with a smaller assigned uncertainty carries greater weight.
Treating different spacings as uncorrelated is a working convention for the weights and their propagation.

Each fit $j$ uses one fitting function and one spacing range, with $n_j$ input masses and $k_j$ parameters.
Its minimum weighted sum of squared residuals is $\chi_j^2$.
We now propagate the input uncertainties $\sigma_a$ to the uncertainty $\sigma_j$ of its continuum intercept.
Let $\mathcal S_j$ be the set of included spacings, and collect their masses $m_*(a)$ in a column vector $\boldsymbol y_j$.
Order the fitted coefficients as $\boldsymbol c_j=(c_{0,j},c_{2,j},c_{2L,j},c_{4,j},\ldots,c_{2q+2,j})^{\mathsf T}$, so that $k_j=q+3$.
With $z_a=ga$, define the $n_j\times k_j$ matrix $X_j$ by
\begin{equation}
 \begin{aligned}
 (X_j)_{a1}&=1,\qquad (X_j)_{a2}=z_a^2,\qquad
 (X_j)_{a3}=z_a^2\ln z_a,\\
 (X_j)_{a,2+i}&=z_a^{2i},
 \qquad i=2,\ldots,q+1,\quad a\in\mathcal S_j.
 \end{aligned}
 \label{eq:continuum-fit-matrix}
\end{equation}
Each row of $X_j\boldsymbol c_j$ is the fitting function evaluated at one spacing.
Under the stated convention that input uncertainties at different spacings are uncorrelated, the weight matrix and the covariance matrix used for their propagation are
\begin{equation}
 W_j=\operatorname{diag}_{a\in\mathcal S_j}(\sigma_a^{-2}),
 \qquad
 \Sigma_j=\operatorname{diag}_{a\in\mathcal S_j}(\sigma_a^2)=W_j^{-1}.
 \label{eq:continuum-input-covariance}
\end{equation}
In this notation, Eq.~\eqref{eq:continuum-weighted-residuals} with $\mathcal S=\mathcal S_j$ reads $\chi^2(\boldsymbol c)=(\boldsymbol y_j-X_j\boldsymbol c)^{\mathsf T}W_j(\boldsymbol y_j-X_j\boldsymbol c)$, which is minimized by
\begin{equation}
 \boldsymbol c_j=L_j\boldsymbol y_j,\qquad
 L_j=(X_j^{\mathsf T}W_jX_j)^{-1}X_j^{\mathsf T}W_j.
 \label{eq:continuum-fit-response}
\end{equation}
For a fixed fitting function, spacing range and weights, replacing $\boldsymbol y_j$ by $\boldsymbol y_j+\Delta\boldsymbol y_j$ changes the fitted coefficients by $\Delta\boldsymbol c_j=L_j\Delta\boldsymbol y_j$.
Propagating the input uncertainties gives the covariance matrix $C_j$ of the fitted coefficients,
\begin{equation}
 C_j=L_j\Sigma_jL_j^{\mathsf T}
 =(X_j^{\mathsf T}W_jX_j)^{-1}.
 \label{eq:continuum-parameter-covariance}
\end{equation}
Since the continuum intercept $c_{0,j}$ is the first coefficient, its propagated numerical uncertainty is
\begin{equation}
 \sigma_j^2=(C_j)_{11}
 =\bigl[(X_j^{\mathsf T}W_jX_j)^{-1}\bigr]_{11}.
 \label{eq:continuum-intercept-error}
\end{equation}
We use Eq.~\eqref{eq:continuum-intercept-error} directly, without multiplying $\sigma_j$ by $\sqrt{\chi_j^2/(n_j-k_j)}$.
The assigned input uncertainties $\sigma_a$ already determine the magnitude of the propagated uncertainty; the residuals in the continuum fits enter through~$\chi_j^2$ in the retention criterion and the weights below.
We retain fits satisfying
\begin{equation}
 n_j-k_j\ge1,\qquad \frac{\chi_j^2}{n_j-k_j}\le2 .
 \label{eq:fit-criteria}
\end{equation}
The difference $n_j-k_j$ is the number of residual degrees of freedom, abbreviated as $\mathrm{dof}$.
The second condition is a goodness-of-fit criterion at the scale of the assigned uncertainties; since the $\sigma_a$ are systematic estimates rather than widths of a distribution, we attach no significance level to it.

We combine the retained fits by a weighted average.
Each fit receives the weight
\begin{equation}
 w_j=\frac{e^{-\mathrm{AIC}_j/2}}{\sum_{j'}e^{-\mathrm{AIC}_{j'}/2}},\qquad
 \mathrm{AIC}_j=\chi_j^2+2k_j+2N_{{\rm cut},j},
 \label{eq:akaike-weights}
\end{equation}
where the sum runs over the retained fits and $N_{{\rm cut},j}=11-n_j$ is the number of spacings omitted from fit $j$.
The first two terms of $\mathrm{AIC}_j$ form the Akaike information criterion~\cite{Akaike:1974vps} for a Gaussian likelihood with known variances.
The last term is the penalty for omitted data introduced in Ref.~\cite{Jay:2020jkz}.
The continuum estimate and the two parts of its uncertainty are
\begin{equation}
 \langle c_0\rangle=\sum_jw_jc_{0,j},\qquad
 \Delta_{\rm in}^2=\sum_jw_j\sigma_j^2,\qquad
 \Delta_{\rm fit}^2=\sum_jw_j\bigl(c_{0,j}-\langle c_0\rangle\bigr)^2,
 \label{eq:akaike-average}
\end{equation}
and we quote
\begin{equation}
 \Delta_{\rm quoted}=\sqrt{\Delta_{\rm in}^2+\Delta_{\rm fit}^2}
 \label{eq:quoted-error}
\end{equation}
as the total uncertainty in $\langle c_0 \rangle$.
The term $\Delta_{\rm in}$ measures the input uncertainty propagated through the individual fits; $\Delta_{\rm fit}$ measures the spread of their continuum intercepts.
These expressions have the form of the mean and variance used in the model averages of Refs.~\cite{Jay:2020jkz,Neil:2022joj}.

Here the weights provide a rule for combining fits, without a probabilistic interpretation.
The Bayesian derivation in Refs.~\cite{Jay:2020jkz,Neil:2022joj} treats data as sample averages with a Gaussian likelihood and a covariance estimated from the sample.
Our VUMPS calculations are deterministic: $\sigma_a$ estimates systematic effects of the bond-dimension and mass extrapolations, and errors at different spacings may be correlated.
We therefore do not interpret $\chi_j^2$ as minus twice a log-likelihood, $w_j$ as a probability of a fitting function, or $\Delta_{\rm quoted}$ as a posterior standard deviation.

We adopt this rule because it favours small residuals while penalizing extra parameters and omitted spacings.
The uncertainty includes contributions from both imprecise fits and disagreement among fits.
There is no separate precision cut on $\sigma_j$: even a fit with a large propagated uncertainty can enter the average.
Section~\ref{sec:budget} tests changes to these choices, including the overall scale assigned to $\sigma_a$.

Because the fits reuse the same input masses, they are not independent measurements.
We therefore use the weighted root mean square of their uncertainties, $\Delta_{\rm in}$, without dividing by a factor that grows with the number of fits.
For comparison, one could propagate the input uncertainties directly through the average while holding the weights fixed.
Let $\ell_{j,a}=(L_j)_{1a}$ for a spacing included in fit $j$, and zero otherwise.
By Eq.~\eqref{eq:continuum-fit-response}, this coefficient measures how much $c_{0,j}$ changes when $m_*(a)$ changes by one unit, so $\sigma_j^2=\sum_a\ell_{j,a}^2\sigma_a^2$.
At fixed weights the average is $\langle c_0\rangle=\sum_a(\sum_jw_j\ell_{j,a})m_*(a)$.
Propagating uncorrelated input uncertainties through this expression would give
\[
 \left[\sum_a\left(\sum_jw_j\ell_{j,a}\right)^2\sigma_a^2\right]^{1/2}
 \leq\Delta_{\rm in},
\]
where the inequality follows from Jensen's inequality.
We use the larger value $\Delta_{\rm in}$ in the quoted uncertainty.

Holding the weights fixed is only a comparison: in the numerical checks of Sec.~\ref{sec:budget}, we repeat the fits, the retention test and the weighting after changing the input masses.
Combining $\Delta_{\rm in}$ and $\Delta_{\rm fit}$ in quadrature is an empirical prescription. The resulting uncertainty is a systematic estimate, not a confidence interval.

\subsection{Continuum extrapolation}
\label{sec:continuum}

\begin{figure}[htbp]
\centering
\includegraphics[width=\textwidth]{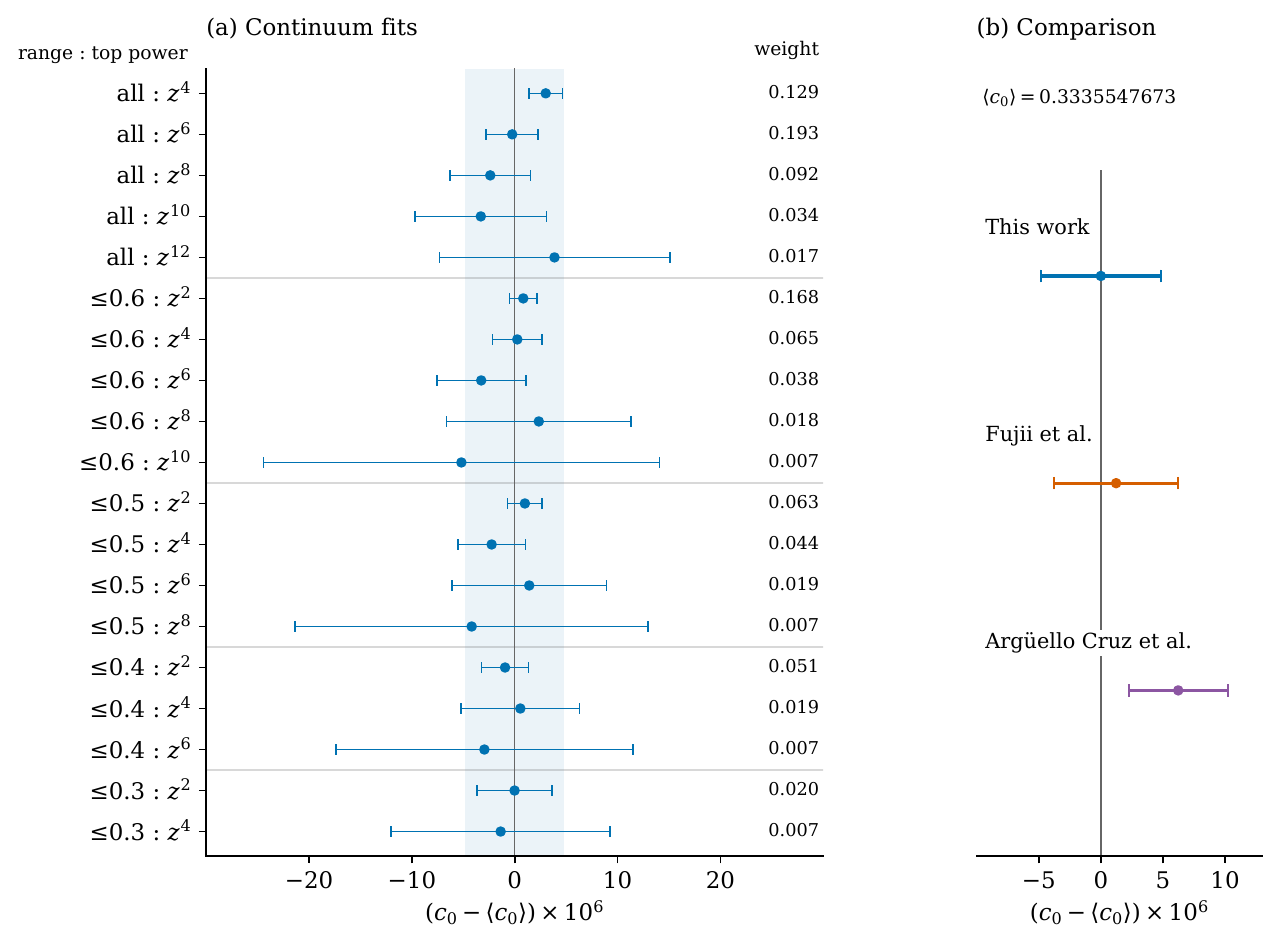}
\caption{Continuum-fit sensitivity and comparison of results.
(a) The nineteen fits satisfying Eq.~\eqref{eq:fit-criteria}, grouped by the upper limit on $ga$.
The row labels give the range and the highest power of $z$ in the fitting function, and the column on the right gives the weight $w_j$.
Error bars show the propagated numerical uncertainties $\sigma_j$ in the continuum intercepts.
Offsets are relative to the unrounded average $\langle c_0\rangle$ in Eq.~\eqref{eq:average-values}, which is marked by the solid vertical line.
The shaded band is $\langle c_0\rangle\pm4.83\times10^{-6}$.
(b) This result and the published values of Refs.~\cite{Fujii:2024giv,ArguelloCruz:2024xzv}, with their quoted uncertainties.
The fits in (a) are correlated analyses of the same data.}
\label{fig:stability}
\end{figure}

The six fitting functions and five ranges give 24 fits with at least one residual degree of freedom.
Nineteen of them satisfy Eq.~\eqref{eq:fit-criteria}.
Table~\ref{tab:fits} lists all 24.
The function with only $c_0$, $c_2$ and $c_{2L}$ gives $\chi_j^2/(n_j-k_j)=3.05$ on all eleven spacings and is retained on the four shorter ranges.
The other four rejected fits have a single residual degree of freedom and $\chi_j^2/(n_j-k_j)$ between $3.24$ and $3.29$.
The retained intercepts lie between $0.333\,549\,6$ and $0.333\,558\,6$, with propagated uncertainties $\sigma_j$ between $1.4\times10^{-6}$ and $19.3\times10^{-6}$.
No single fit dominates the average: the largest weight is $0.193$.
Summed over the ranges, the weights are $0.301$, $0.265$, $0.258$, $0.117$, $0.042$ and $0.017$ for the functions ending at $z^2,z^4,\ldots,z^{12}$.
Summed over the functions, they are $0.466$ for all eleven spacings and $0.297$, $0.133$, $0.077$ and $0.027$ for $z\le0.6$, $0.5$, $0.4$ and $0.3$.
Equations~\eqref{eq:akaike-average} and~\eqref{eq:quoted-error} give
\begin{equation}
 \langle c_0\rangle=0.333\,554\,77,\qquad
 \Delta_{\rm in}=4.42\times10^{-6},\qquad
 \Delta_{\rm fit}=1.93\times10^{-6},\qquad
 \Delta_{\rm quoted}=4.83\times10^{-6}.
 \label{eq:average-values}
\end{equation}
The fixed-weight propagation described in Sec.~\ref{sec:contmodel} would give $2.4\times10^{-6}$ in place of $\Delta_{\rm in}$.
The difference comes from fits with small weights and large $\sigma_j$: the seven retained fits with $\sigma_j>7\times10^{-6}$ carry $8\%$ of the weight and contribute $61\%$ of $\Delta_{\rm in}^2$.
The same weights give $\langle c_{2L}\rangle=-1.00\times10^{-3}$, with an uncertainty of $0.60\times10^{-3}$ defined as in Eqs.~\eqref{eq:akaike-average} and~\eqref{eq:quoted-error}.
The mass at the finest spacing differs from $\langle c_0\rangle$ by $7.82\times10^{-5}$, about $2\%$ of the change of $m_*(a)$ across the eleven spacings.
Figure~\ref{fig:stability} shows the retained estimates and the comparison with staggered-fermion results.

\begin{table}[htbp]
\centering
\small
\begin{tabular}{llcccccrrc}
\toprule
Range & Highest power & $n_j-k_j$ & $c_{0,j}$ & $\sigma_j\,[10^{-6}]$ & $\chi_j^2/(n_j-k_j)$ & $\mathrm{AIC}_j$ & $c_{2L,j}$ & $\sigma(c_{2L,j})$ & $w_j$\\
\midrule
All & $z^{2}$ & 8 & 0.3335528 & 1.07 & 3.05 & 30.36 & $-10.09$ & 0.11 & -- \\
All & $z^{4}$ & 7 & 0.3335578 & 1.62 & 1.08 & 15.57 & $-8.15$ & 0.49 & 0.129 \\
All & $z^{6}$ & 6 & 0.3335545 & 2.53 & 0.79 & 14.76 & $-10.18$ & 1.30 & 0.193 \\
All & $z^{8}$ & 5 & 0.3335524 & 3.93 & 0.85 & 16.25 & $-12.17$ & 3.08 & 0.092 \\
All & $z^{10}$ & 4 & 0.3335515 & 6.39 & 1.05 & 18.22 & $-13.30$ & 6.94 & 0.034 \\
All & $z^{12}$ & 3 & 0.3335586 & 11.20 & 1.20 & 19.61 & $-2.09$ & 15.99 & 0.017 \\
\midrule
$z\le0.6$ & $z^{2}$ & 6 & 0.3335556 & 1.36 & 0.84 & 15.04 & $-9.36$ & 0.26 & 0.168 \\
$z\le0.6$ & $z^{4}$ & 5 & 0.3335550 & 2.42 & 0.99 & 16.95 & $-9.68$ & 1.13 & 0.065 \\
$z\le0.6$ & $z^{6}$ & 4 & 0.3335515 & 4.32 & 1.00 & 17.99 & $-12.96$ & 3.54 & 0.038 \\
$z\le0.6$ & $z^{8}$ & 3 & 0.3335571 & 8.97 & 1.16 & 19.49 & $-5.42$ & 11.16 & 0.018 \\
$z\le0.6$ & $z^{10}$ & 2 & 0.3335496 & 19.25 & 1.65 & 21.29 & $-19.83$ & 34.46 & 0.007 \\
$z\le0.6$ & $z^{12}$ & 1 & 0.3335548 & 42.51 & 3.27 & 23.27 & $-5.84$ & 107.61 & -- \\
\midrule
$z\le0.5$ & $z^{2}$ & 5 & 0.3335558 & 1.68 & 1.00 & 17.01 & $-9.31$ & 0.44 & 0.063 \\
$z\le0.5$ & $z^{4}$ & 4 & 0.3335525 & 3.30 & 0.93 & 17.71 & $-11.61$ & 2.07 & 0.044 \\
$z\le0.5$ & $z^{6}$ & 3 & 0.3335562 & 7.53 & 1.14 & 19.42 & $-7.24$ & 8.33 & 0.019 \\
$z\le0.5$ & $z^{8}$ & 2 & 0.3335506 & 17.17 & 1.64 & 21.29 & $-17.21$ & 28.70 & 0.007 \\
$z\le0.5$ & $z^{10}$ & 1 & 0.3335518 & 37.73 & 3.29 & 23.29 & $-13.97$ & 90.80 & -- \\
\midrule
$z\le0.4$ & $z^{2}$ & 4 & 0.3335538 & 2.28 & 0.86 & 17.44 & $-10.22$ & 0.85 & 0.051 \\
$z\le0.4$ & $z^{4}$ & 3 & 0.3335553 & 5.77 & 1.12 & 19.36 & $-8.77$ & 5.23 & 0.019 \\
$z\le0.4$ & $z^{6}$ & 2 & 0.3335518 & 14.43 & 1.65 & 21.29 & $-14.33$ & 21.65 & 0.007 \\
$z\le0.4$ & $z^{8}$ & 1 & 0.3335491 & 31.94 & 3.28 & 23.28 & $-20.74$ & 71.12 & -- \\
\midrule
$z\le0.3$ & $z^{2}$ & 3 & 0.3335548 & 3.64 & 1.11 & 19.34 & $-9.59$ & 2.10 & 0.020 \\
$z\le0.3$ & $z^{4}$ & 2 & 0.3335534 & 10.64 & 1.66 & 21.32 & $-11.34$ & 13.09 & 0.007 \\
$z\le0.3$ & $z^{6}$ & 1 & 0.3335472 & 24.96 & 3.24 & 23.24 & $-24.24$ & 48.61 & -- \\
\bottomrule
\end{tabular}
\caption{The 24 fits of Eqs.~\eqref{eq:family}--\eqref{eq:family-additional} with at least one residual degree of freedom.
The second column gives the highest power of $z$ in the fitting function.
The fitted coefficient $c_{2L,j}$ and its propagated uncertainty are in units of $10^{-4}$.
A dash in the last column marks a fit that fails Eq.~\eqref{eq:fit-criteria} and does not enter the average.
The weights $w_j$ are those of Eq.~\eqref{eq:akaike-weights}.}
\label{tab:fits}
\end{table}

The primary result is
\begin{equation}
 (m/g)_c=\mcvalerr.
 \label{eq:result}
\end{equation}
The unrounded uncertainty from Eq.~\eqref{eq:quoted-error} is $4.83\times10^{-6}$.
This result uses a free leading logarithmic coefficient, the six polynomial truncations in Eq.~\eqref{eq:family-additional} and the five ranges, and is subject to Eq.~\eqref{eq:fit-criteria} and the weights of Eq.~\eqref{eq:akaike-weights}.
It agrees within the quoted uncertainties with both recent staggered-fermion determinations~\cite{Fujii:2024giv,ArguelloCruz:2024xzv}.
It lies $1.2\times10^{-6}$ below the value $0.333\,556(5)$ of Ref.~\cite{Fujii:2024giv} and $6.2\times10^{-6}$ below the value $0.333\,561(4)$ of Ref.~\cite{ArguelloCruz:2024xzv}.
The second difference is comparable to the two uncertainties added in quadrature: $6.3\times10^{-6}$ using our unrounded uncertainty and $6.4\times10^{-6}$ using its rounded value.

\subsection{Logarithms at order $z^4$ and other fitting functions}
\label{sec:logcheck}

The primary fitting functions include $z^2\ln z$ with a free coefficient and even powers through $z^{12}$, but no higher-order logarithms.
We test these choices by changing the highest power and the logarithmic terms.
We then explain why logarithms at order $z^4$ are omitted from the primary analysis and what assumption this requires.
These tests are reported separately; they do not enter the quoted uncertainty.

\medskip
\noindent\textbf{Highest power.}
Table~\ref{tab:truncation} shows the average when the family of Eq.~\eqref{eq:family-additional} stops at a lower or a higher power.
As the highest power grows from $z^4$ to $z^{16}$, the average changes from $0.333\,555\,58$ to $0.333\,554\,72$, that is, by $-0.9\times10^{-6}$.
The quoted uncertainty grows from $3.0\times10^{-6}$ to $5.2\times10^{-6}$.
The growth comes from $\Delta_{\rm in}$: each added power admits fits with larger $\sigma_j$, and Eq.~\eqref{eq:fit-criteria} places no bound on $\sigma_j$.
Extending the family to $z^{14}$ adds one retained fit and raises the quoted uncertainty from $4.8\times10^{-6}$ to $5.2\times10^{-6}$.
Extending it further to $z^{16}$ adds no retained fit.
Thus the cutoff at $z^{12}$ is a fitting choice, but extending the family through $z^{16}$ leaves the rounded result unchanged.

\medskip
\noindent\textbf{Other families.}
Table~\ref{tab:families} gives the averages for other families of fitting functions.
Each uses the five ranges, Eq.~\eqref{eq:fit-criteria} and Eq.~\eqref{eq:akaike-weights}, with $k_j$ counting every free coefficient.

Setting $c_{2L}=0$ raises the average by $6.4\times10^{-6}$, more than the primary quoted uncertainty.
Two features of the fits favour including this logarithm.
First, the smallest $\mathrm{AIC}_j$ is $14.8$ in the primary family, compared with $17.6$ in the family without the logarithm.
Second, the primary functions ending at $z^4$ and $z^6$, fitted to all eleven spacings, give values of $c_{2L}$ whose magnitudes are many times their propagated uncertainties (Table~\ref{tab:fits}).
In a single average over the union of the two families, the functions with the logarithm carry $80\%$ of the weight, and the average is $0.333\,556\,02$ with a quoted uncertainty of $5.2\times10^{-6}$, higher than the primary value by $1.3\times10^{-6}$.
The primary result thus depends on including the logarithm at order $z^2$. This term is allowed by the matching power count of Sec.~\ref{sec:c2}, but the normal-ordering factor alone does not establish its coefficient in the critical mass.

The next three families include logarithms at order $z^4$.
The first adds $c_{4L}z^4\ln z$, and the second adds $c_{4L}z^4\ln z+c_{4LL}z^4(\ln z)^2$, to every primary function that contains $z^4$.
The third replaces $R_q$ by $z^4\sum_{k=0}^{p}d_k(\ln z)^k$ with free coefficients $d_k$ and $p=0,\ldots,3$, without higher powers of $z$.
Their averages differ from the primary value by $-0.9\times10^{-6}$, $-0.4\times10^{-6}$ and $-0.7\times10^{-6}$.
Their quoted uncertainties are $8.1\times10^{-6}$, $14.1\times10^{-6}$ and $10.1\times10^{-6}$.
The last family adds $c_{2LL}z^2(\ln z)^2$ to every primary function.
Its average is lower by $3.3\times10^{-6}$, with a quoted uncertainty of $15.9\times10^{-6}$.
All four shifts are smaller than the primary quoted uncertainty, and all four quoted uncertainties are larger than it by factors between $1.7$ and $3.3$.

\medskip
\noindent\textbf{Why the primary family omits logarithms at order $z^4$.}
The power counting in Sec.~\ref{sec:physicalcounting} allows terms $z^4(\ln z)^k$.
Including these terms shifts the average by less than $1\times10^{-6}$, but increases its uncertainty by factors of $1.7$ to $2.9$.
The reason is that, over the measured spacing range, these logarithmic terms closely resemble combinations of the primary fitting functions. The data have difficulty distinguishing their contributions.
We quantify this similarity and its effect on the continuum intercept in two tests.

First, over the range $0.1\le z\le0.8$ of the data, a term $z^4(\ln z)^k$ is close to a linear combination of functions that the primary family already contains.
To measure this similarity, we use the eleven values of $z^4(\ln z)^k$ as test data and fit them with a primary function, using weights $1/\sigma_a^2$.
We divide the weighted root mean square of the residuals by the weighted root mean square of the test data. A small ratio means that the primary function closely reproduces the logarithmic term over these spacings.
For the function ending at $z^6$ the ratio is $0.19\%$, $0.15\%$ and $1.8\%$ for $k=1,2,3$.
For the function ending at $z^8$ it is $0.04\%$, $0.10\%$ and $0.3\%$.
A logarithmic term at order $z^4$ therefore changes the fitted coefficients of the even powers and leaves only a small residual.

Second, we add $b\,z^4(\ln z)^k$ to the eleven masses $m_*(a)$ and repeat the complete procedure, including the retention test and the weights.
The added term vanishes at $z=0$, so it leaves the underlying continuum value unchanged.
Any change in the fitted $\langle c_0\rangle$ therefore measures the sensitivity of our extrapolation to an omitted term of this form and size.
Table~\ref{tab:injection} gives the shifts for $k=1,2,3$ and $|b|$ between $10^{-3}$ and $4\times10^{-3}$.
For comparison, the primary functions ending at $z^4$, $z^6$ and $z^8$ give, on all eleven spacings, $c_2$ between $5.2\times10^{-3}$ and $5.7\times10^{-3}$, $c_{2L}$ between $-1.2\times10^{-3}$ and $-0.8\times10^{-3}$, and $c_4$ between $-0.2\times10^{-3}$ and $1.3\times10^{-3}$.
The added terms change individual masses by up to $3.7\times10^{-4}$, about $300$ times the assigned uncertainties.
The average moves by at most $5.2\times10^{-6}$.
For $|b|=10^{-3}$ the shifts lie between $-2.8\times10^{-6}$ and $+0.5\times10^{-6}$.
For $|b|\le4\times10^{-3}$, sixteen of the eighteen shifts are smaller than the quoted uncertainty of $4.8\times10^{-6}$; the two exceptions are $-5.2\times10^{-6}$ and $-4.9\times10^{-6}$, both for $k=3$ with negative $b$.
The shift is not proportional to $b$, because the weights and the set of retained fits change with the data.

The larger uncertainties in the families with order-$z^4$ logarithms arise because the data constrain their coefficients poorly; they are not evidence that these terms are large.
For example, set $p=3$ in the family $z^4\sum_{k=0}^{p}d_k(\ln z)^k$, so that all four coefficients at order $z^4$ are free.
The fit to all eleven spacings gives $c_{4L}=0.009(29)$, $c_{4LL}=-0.005(17)$ and $c_{4LLL}=0.006(16)$, together with $c_0=0.333\,558$ and $\sigma_j=20\times10^{-6}$.
The coefficient uncertainties are an order of magnitude larger than the fitted $|c_{2L}|$.
Moreover, $c_0$ is strongly correlated with all six other parameters: the magnitudes of the correlation coefficients range from $0.93$ to $0.98$.
The logarithmic coefficients can therefore change together by about $2\times10^{-2}$ while largely compensating one another over the measured range.
Such a change can move $c_0$ by $2\times10^{-5}$ while raising $\chi^2$ by only one unit.
With one logarithm at order $z^4$, $p=1$, the same fit gives $c_{4L}=-0.9(5)\times10^{-3}$.

The primary uncertainty therefore relies on an assumption about coefficient sizes: the coefficients of $z^4(\ln z)^k$ are not much larger than a few times $10^{-3}$.
This is the scale of the coefficients determined by the data for $z^2\ln z$ and $z^4$.
For the terms and sizes tested in Table~\ref{tab:injection}, the shifts are comparable to or smaller than the quoted uncertainty.
If the order-$z^4$ logarithmic coefficients are left unrestricted, the corresponding fitting families instead give uncertainties of order $10^{-5}$.
We have not calculated these coefficients, and their assumed size does not follow from power counting.
\begin{table}[htbp]
\centering
\small
\begin{tabular}{lccccc}
\toprule
Highest power in the family & Fits & $\langle c_0\rangle$ & $\Delta_{\rm in}$ & $\Delta_{\rm fit}$ & $\Delta_{\rm quoted}$\\
\midrule
$z^{4}$ & 9 & 0.33355558 & 2.55 & 1.50 & 2.95\\
$z^{6}$ & 13 & 0.33355512 & 3.16 & 1.58 & 3.53\\
$z^{8}$ & 16 & 0.33355486 & 3.76 & 1.76 & 4.15\\
$z^{10}$ & 18 & 0.33355470 & 4.21 & 1.88 & 4.61\\
$z^{12}$ (primary) & 19 & 0.33355477 & 4.42 & 1.93 & 4.83\\
$z^{14}$ & 20 & 0.33355472 & 4.80 & 2.01 & 5.20\\
$z^{16}$ & 20 & 0.33355472 & 4.80 & 2.01 & 5.20\\
\bottomrule
\end{tabular}
\caption{Averages when the family of Eq.~\eqref{eq:family-additional} is truncated at a lower or a higher power of $z$.
Uncertainties are in units of $10^{-6}$.
The column ``Fits'' gives the number of retained fits.}
\label{tab:truncation}
\end{table}

\begin{table}[htbp]
\centering
\small
\begin{tabular}{p{6.6cm}cccccc}
\toprule
Family & Fits & $\langle c_0\rangle$ & $\Delta_{\rm in}$ & $\Delta_{\rm fit}$ & $\Delta_{\rm quoted}$ & Shift\\
\midrule
Primary, Eqs.~\eqref{eq:family}--\eqref{eq:family-additional} & 19 of 24 & 0.33355477 & 4.42 & 1.93 & 4.83 & --\\
No logarithm, $c_{2L}=0$ & 13 of 27 & 0.33356113 & 2.93 & 1.66 & 3.37 & $+6.36$\\
$c_{4L}z^4\ln z$ added & 15 of 20 & 0.33355382 & 7.65 & 2.73 & 8.12 & $-0.94$\\
$c_{4L}z^4\ln z+c_{4LL}z^4(\ln z)^2$ added & 11 of 16 & 0.33355432 & 13.24 & 4.76 & 14.07 & $-0.45$\\
$R_q$ replaced by $z^4\sum_{k=0}^{p}d_k(\ln z)^k$, $p\le3$ & 14 of 17 & 0.33355405 & 9.58 & 3.33 & 10.14 & $-0.72$\\
$c_{2LL}z^2(\ln z)^2$ added & 16 of 20 & 0.33355149 & 14.93 & 5.45 & 15.90 & $-3.28$\\
\bottomrule
\end{tabular}
\caption{Averages for other families of fitting functions, with the ranges, retention criterion and weights of the primary analysis.
The column ``Fits'' gives the number of retained fits and the number of fits with at least one residual degree of freedom.
The terms at order $z^4$ are added to the primary functions that contain $z^4$; the term $c_{2LL}z^2(\ln z)^2$ is added to all six primary fitting functions.
Uncertainties and shifts of $\langle c_0\rangle$ relative to the primary value are in units of $10^{-6}$.}
\label{tab:families}
\end{table}

\begin{table}[htbp]
\centering
\small
\begin{tabular}{rcccccc}
\toprule
 & \multicolumn{2}{c}{$b\,z^4\ln z$} & \multicolumn{2}{c}{$b\,z^4(\ln z)^2$} & \multicolumn{2}{c}{$b\,z^4(\ln z)^3$}\\
\cmidrule(lr){2-3}\cmidrule(lr){4-5}\cmidrule(lr){6-7}
$b\,[10^{-3}]$ & Shift & $\Delta_{\rm quoted}$ & Shift & $\Delta_{\rm quoted}$ & Shift & $\Delta_{\rm quoted}$\\
\midrule
$-4$ & $+1.73$ & 6.11 & $-3.00$ & 6.06 & $-4.89$ & 7.39\\
$-2$ & $+0.68$ & 5.80 & $-1.89$ & 5.79 & $-5.25$ & 4.57\\
$-1$ & $+0.04$ & 5.37 & $-1.19$ & 5.71 & $-2.77$ & 3.52\\
$+1$ & $-1.61$ & 4.10 & $-0.18$ & 4.25 & $+0.53$ & 6.08\\
$+2$ & $-2.96$ & 5.27 & $-0.60$ & 5.21 & $+0.53$ & 6.89\\
$+4$ & $-4.46$ & 5.70 & $+1.02$ & 5.92 & $+0.63$ & 7.69\\
\bottomrule
\end{tabular}
\caption{Shift of the average $\langle c_0\rangle$ when a logarithmic term of order $z^4$ with coefficient $b$ is added to the eleven masses $m_*(a)$ and the complete procedure is repeated.
Shifts and the resulting quoted uncertainties are in units of $10^{-6}$.
For $|b|=10^{-3}$ the added term changes individual masses by up to $9.1\times10^{-5}$, $3.4\times10^{-5}$ and $2.1\times10^{-5}$ for the first, second and third power of $\ln z$; these changes scale with $|b|$.
The number of retained fits is between 14 and 20.}
\label{tab:injection}
\end{table}

\FloatBarrier
\subsection{Uncertainty budget and numerical checks}
\label{sec:budget}

Table~\ref{tab:budget} gives the two contributions used in Eq.~\eqref{eq:quoted-error}.
The shifts caused by varying the fitted bond dimensions and allowing quadratic mass dependence in the joint fit are included in \(\Delta_{\rm in}\).
Propagating each of the three terms in Eq.~\eqref{eq:input-error} separately gives contributions of $3.01\times10^{-6}$ from the least-squares residuals, $3.12\times10^{-6}$ from the bond-dimension variation, and $0.91\times10^{-6}$ from the quadratic mass dependence.
These contributions add in quadrature to $\Delta_{\rm in}$.

\begin{table}[htbp]
\centering
\begin{tabular}{lcl}
\toprule
Source & Size $[10^{-6}]$ & Prescription \\
\midrule
Propagated input uncertainty $\Delta_{\rm in}$ & 4.42 & Eq.~\eqref{eq:akaike-average} \\
Variation among fits $\Delta_{\rm fit}$ & 1.93 & Eq.~\eqref{eq:akaike-average} \\
\midrule
Combined uncertainty $\Delta_{\rm quoted}$ & 4.83 & Quadrature, Eq.~\eqref{eq:quoted-error} \\
\bottomrule
\end{tabular}
\caption{Uncertainty prescription for the primary result, Eq.~\eqref{eq:result}.
The sensitivity to logarithms at order $z^4$, to other fitting functions and to other finite-$D$ extrapolations is reported separately.}
\label{tab:budget}
\end{table}

\medskip
\noindent\textbf{Averaging prescription.}
Table~\ref{tab:prescription-checks} shows the effect of changing elements of the averaging prescription with the input masses fixed.
Removing the condition on $\chi_j^2/(n_j-k_j)$ changes the average by $-0.04\times10^{-6}$ and raises the quoted uncertainty to $6.0\times10^{-6}$.
The four fits with a single residual degree of freedom then enter with weight $0.003$ each and $\sigma_j$ between $25\times10^{-6}$ and $43\times10^{-6}$.
Alternatively, we can remove the residual-size condition and require at least two residual degrees of freedom, $n_j-k_j\ge2$.
This gives the same result to the reported precision: it retains the primary fits and adds only one fit, with weight below $10^{-3}$.
Removing the penalty for omitted spacings changes the average by $-0.4\times10^{-6}$, and using only the fits to all eleven spacings changes it by $+0.2\times10^{-6}$.
The weights depend on the overall scale of the $\sigma_a$ through $\chi_j^2$.
Doubling every $\sigma_a$ moves the average by $+0.5\times10^{-6}$ and gives a quoted uncertainty of $8.5\times10^{-6}$; all 24 fits are then retained.
Halving every $\sigma_a$ leaves no fit that satisfies Eq.~\eqref{eq:fit-criteria}; without that condition the average moves by $-0.9\times10^{-6}$.
The $\sigma_a$ are systematic estimates whose overall scale is not fixed by a statistical argument, so this dependence is a limitation of the prescription.

\begin{table}[htbp]
\centering
\small
\begin{tabular}{p{8.2cm}ccc}
\toprule
Change of the prescription & Shift of $\langle c_0\rangle$ & $\Delta_{\rm quoted}$ & Fits \\
\midrule
No condition on $\chi_j^2/(n_j-k_j)$ & $-0.04$ & 6.04 & 24 \\
$n_j-k_j\ge2$ in place of the condition on $\chi_j^2/(n_j-k_j)$ & $0.00$ & 4.83 & 20 \\
No penalty for omitted spacings & $-0.39$ & 5.73 & 19 \\
Fits to all eleven spacings only & $+0.16$ & 4.36 & 5 \\
Every $\sigma_a$ multiplied by 2 & $+0.47$ & 8.53 & 24 \\
Every $\sigma_a$ multiplied by $1/2$ & -- & -- & 0 \\
Every $\sigma_a$ multiplied by $1/2$, no condition on $\chi_j^2/(n_j-k_j)$ & $-0.92$ & 5.44 & 24 \\
\bottomrule
\end{tabular}
\caption{Changes of the average $\langle c_0\rangle$ under changes of the averaging prescription, with the primary input masses.
Shifts and the resulting quoted uncertainties are in units of $10^{-6}$.
The column ``Fits'' gives the number of retained fits.}
\label{tab:prescription-checks}
\end{table}

\medskip
\noindent\textbf{Changes to the input masses.}
Table~\ref{tab:continuum-checks} shows how alternative determinations of the lattice critical masses change the continuum result. For each row, we use the revised estimates \(m_*(a)\) at the spacings where the alternative calculation was performed and keep the original estimates at the remaining spacings. In most rows, we change only these mass estimates: their assigned uncertainties \(\sigma_a\) remain those used in the primary analysis. The row labeled “Use the extended scaling fit and its own input uncertainties” changes both the masses and their uncertainties. For every row, we repeat the continuum fits, apply the fit-selection criteria, and recalculate the Akaike weights.

\begin{table}[htbp]
\centering
\small
\begin{tabular}{p{8.2cm}ccc}
\toprule
Change from the primary analysis & Shift of $\langle c_0\rangle$ & $\Delta_{\rm quoted}$ & Fits \\
\midrule
Omit $D=160$ where available & $-0.30$ & 5.02 & 19 \\
Omit $D=320$ where available & $-1.61$ & 6.47 & 23 \\
Allow quadratic mass dependence of the intercept & $-0.73$ & 5.83 & 19 \\
Use the extended scaling fit, Eq.~\eqref{eq:extfit} & $+0.26$ & 5.03 & 19 \\
Use the extended scaling fit and its own input uncertainties & $-0.45$ & 6.60 & 19 \\
Apply the mass shifts of Eq.~\eqref{eq:local-cutoff-shift} at four spacings & $+0.05$ & 4.78 & 19 \\
Use the two-stage masses of Table~\ref{tab:seqcmp} at five spacings & $-3.18$ & 9.40 & 24 \\
\bottomrule
\end{tabular}
\caption{Changes of the average $\langle c_0\rangle$ under changes of the input masses, with the fitting functions, ranges, retention criterion and weights of the primary analysis.
Shifts and the resulting quoted uncertainties are in units of $10^{-6}$, and the column ``Fits'' gives the number of retained fits.
These are sensitivity checks, not additional independent entries in Table~\ref{tab:budget}.}
\label{tab:continuum-checks}
\end{table}

The extended fit, Eq.~\eqref{eq:extfit}, includes quadratic and cubic terms in the signed scaling variable $x=t/\delta$.
Fitting this form simultaneously to the transfer-matrix data at all eleven lattice spacings, with common coefficients $\phi_2$ and $\phi_3$ and separate parameters $m_*(a)$, $p_1$, and $c_1$ at each spacing, gives $\phi_2=-0.025(2)$ and $\phi_3=0.006(2)$.
The correlation coefficient between these two fitted parameters is $-0.93$.
They have the same signs and similar magnitudes as $-0.028(3)$ and $0.008(2)$ obtained from the particular cubic description of the transverse-field Ising-chain data used in Appendix~\ref{sec:tfim}.
Because the coefficients change with the interpolation used for the Ising data, this comparison is qualitative.

To test whether these terms improve predictions at larger bond dimension, we exclude the $D=320$ points and fit both the leading and extended forms using $128\le D\le256$.
For each excluded point $i$, let $\widehat\epsilon_{1,i}$ denote the fitted function evaluated at that point's mass $m_i$ and computed value $\delta_i=\epsilon_{2,i}-\epsilon_{1,i}$.
The $D=320$ calculation thus still supplies $\delta_i$; the fit predicts $\epsilon_1$ at that value of $\delta$.
There are 20 excluded points, five at each of $ga=0.2,0.15,0.125,0.1$.
Their unweighted root-mean-square (rms) deviation is $\mathrm{rms}=[(1/20)\sum_{i=1}^{20}(\epsilon_{1,i}-\widehat\epsilon_{1,i})^2]^{1/2}$.
It is $4.3\times10^{-7}$ for the extended fit and $5.4\times10^{-7}$ for the leading joint fit.
These deviations are in the dimensionless inverse correlation length $\epsilon_1$, not uncertainties in the critical mass.

Using the input uncertainties obtained with the extended fit, the complete continuum procedure gives $0.33355431$ with quoted uncertainty $6.60\times10^{-6}$.

As an exploratory check, we add a term proportional to $\delta^2$ to Eq.~\eqref{eq:jointfit} at the six spacings with sufficient bond-dimension data.
Its coefficient is poorly constrained, and the resulting continuum average is dominated by fits with only one residual degree of freedom.
We therefore do not use this extension to estimate a shift or an additional uncertainty.

\medskip
\noindent\textbf{Local cutoff and solver tolerance.}
We test the local cutoff by increasing $n_{\max}$ from 30 to 34 at $ga=(0.3,0.6,0.7,0.8)$, with $D=192$ and one fixed mass near the transition at each spacing.
Write $\Delta X=X_{34}-X_{30}$ for the difference between the two cutoff values.
Holding the slopes $p_1$ and $c_1$ of the $n_{\max}=30$ joint fit fixed, Eq.~\eqref{eq:jointfit} converts the measured changes in $\epsilon_1$ and $\delta$ into an estimated mass shift,
\begin{equation}
 \Delta m_*(a)\simeq
 -\frac{\Delta\epsilon_1-c_1\Delta\delta}{p_1}.
 \label{eq:local-cutoff-shift}
\end{equation}
Using the joint-fit slopes employed in the present analysis gives $\Delta m_*(a)=(-1.09,-2.89,-3.27,-3.46)\times10^{-7}$ at these four spacings.
These estimates assume that the cutoff dependence can be represented locally by a shift of the fitted critical mass; we have not repeated the full mass and bond-dimension extrapolation at $n_{\max}=34$.
The occupation probability beyond $n=30$ is at most $2\times10^{-9}$ in these checks.
The Zenodo release~\cite{Okuda:2026bosonizeddata} includes the paired transfer-matrix data and fitted slopes, together with a code that reproduces Eq.~\eqref{eq:local-cutoff-shift}.
These mass-shift estimates are reported as cutoff checks and are not separately propagated into the continuum uncertainty.
Applying them to the four masses changes the average by $+0.05\times10^{-6}$ (Table~\ref{tab:continuum-checks}).
Reducing the solver tolerance from $3\times10^{-10}$ to $10^{-10}$ at $ga=0.2$, $D=192,256$, and at $ga=0.125,0.1$, $D=256$, gives $|\Delta\epsilon_1|\le3.8\times10^{-12}$ and $|\Delta\delta|\le9.5\times10^{-11}$.
These changes in $\epsilon_1$ and $\delta$ are much smaller than those associated with the extrapolations.

\medskip
\noindent\textbf{Two-stage extrapolation.}
Table~\ref{tab:seqcmp} compares the joint-fit masses with the two-stage method of Sec.~\ref{sec:delta}: extrapolate in $\delta$ at each fixed mass, then find the mass where the intercept vanishes.
The table gives the difference between the two methods at each lattice spacing.
Replacing the five primary masses by the two-stage values while keeping the primary $\sigma_a$ changes the average by $-3.2\times10^{-6}$, and the quoted uncertainty of that analysis is $9.4\times10^{-6}$ (Table~\ref{tab:continuum-checks}).
We report this as a sensitivity check and do not add it to the quoted uncertainty.

\begin{table}[htb]
\centering
\caption{Two-stage estimates, from separate fixed-mass extrapolations,
compared with the joint-fit values of Table~\ref{tab:perspacing}. Both
use $D\ge128$, with $D=160$ and $D=320$ where available; the two-stage
method uses the five fixed masses at each spacing. Two-stage
uncertainties come from weighted zero-crossing fits with
$\chi^2/\mathrm{dof}$ rescaling. Including $D=96$ changes a two-stage
estimate by at most $6.1\times10^{-6}$. Differences are computed before
rounding.}
\label{tab:seqcmp}
\begin{tabular}{lcccc}
\toprule
$ga$ & two-stage $m_*$ & $\sigma\,[10^{-6}]$ & joint $m_*$ & difference $[10^{-6}]$ \\
\midrule
0.5 & $0.3351155$ & $1.5$ & $0.3351145$ & $+1.0$ \\
0.4 & $0.3345888$ & $0.3$ & $0.3345849$ & $+3.8$ \\
0.3 & $0.3341637$ & $1.5$ & $0.3341608$ & $+2.9$ \\
0.2 & $0.3338427$ & $2.6$ & $0.3338387$ & $+4.0$ \\
0.1 & $0.3336321$ & $1.3$ & $0.3336330$ & $-0.9$ \\
\bottomrule
\end{tabular}
\end{table}

\FloatBarrier

\widowpenalty=10000
\section{Lattice corrections and the fitting assumptions}
\label{sec:c2}

The continuum fits of Sec.~\ref{sec:contmodel} use even powers of $ga$ and a logarithm at order $(ga)^2$.
The argument has two steps. First, a local expansion of lattice corrections determines the allowed powers of $a$ in the effective action.
Second, we assume that the critical mass has a Taylor expansion in the coefficients of these corrections.
Together, these steps give even powers of $ga$ with possible logarithmic factors, but do not determine the logarithmic coefficients.
This section presents the power counting; Appendix~\ref{sec:freecorrelatormatching} gives the matching equations and subtraction procedure.

\subsection{Local corrections to the continuum theory}
\label{sec:localmatching}

The lattice changes propagation at distances comparable to $a$.
At much larger distances, Symanzik's description represents this difference by local insertions in the continuum theory~\cite{Symanzik:1983dc,Symanzik:1983gh,Weisz:2010rg}:
\begin{equation}
 S_{\rm eff}=S_{\rm cont}+\sum_i h_i(a,q)\int d^2X\,\mathcal O_i^R(X;q),
 \qquad X=(x,\tau).
 \label{eq:symanzik}
\end{equation}
An equivalent effective-Hamiltonian expansion organizes the corrections directly in powers of the spatial lattice spacing and relates them to the anisotropic effective action~\cite{Steffens:2026symanzik}.
For matching correlation functions, we use the Euclidean-action form above, with continuous imaginary time and a spatial lattice cutoff.
The index $i$ labels local operators $\mathcal O_i^R$; the superscript $R$ indicates that they are renormalized at the fixed physical scale $q$.
Their coefficients $h_i(a,q)$ are chosen so that continuum correlation functions, including these insertions, reproduce the long-distance lattice correlation functions order by order in $a$.

The simplest example is the free inverse propagator, with $\mu=g/\sqrt\pi$, spatial momentum $k$ and Euclidean frequency $\omega$:
\begin{equation}
 \omega^2+\mu^2+\frac4{a^2}\sin^2\frac{ak}{2}
 =\omega^2+\mu^2+k^2-\frac{a^2k^4}{12}+O(a^4k^6).
 \label{eq:free-matching}
\end{equation}
The first difference is reproduced by
\begin{equation}
 \delta S=-\frac{a^2}{24}\int d^2X\,(\partial_x^2\phi)^2.
 \label{eq:dispersion-correction}
\end{equation}
Interactions generate further coefficients $h_i$.
To determine them, one compares lattice and continuum correlation functions at fixed physical separations, subtracts effects already accounted for by known coefficients, and represents the remaining local difference by the operators in Eq.~\eqref{eq:symanzik}.
The required perturbative correlation functions can be evaluated in the free massless non-compact boson theory.
For the continuum fits, we first need the powers of $a$ that this procedure can produce.

\subsection{Why even powers of the lattice spacing occur}
\label{sec:physicalcounting}

The coefficients $h_i$ describe how propagation and interactions at distances of order $a$ affect correlation functions measured at much larger distances.
To count their possible powers of $a$, expand the cosine interaction and the quadratic potential in correlation functions of the free massless scalar.
The free propagator is logarithmic at short distances.
Consequently, a normal-ordered vertex of charge $k$, proportional to $e^{ik\beta\phi}$ with $\beta=2\sqrt\pi$, has ultraviolet dimension $k^2$, while undifferentiated factors of $\phi$ add logarithms without adding an algebraic power of distance.
Here the vertex charge $k$ is the integer multiplying $\beta\phi$ in the exponential; it is distinct from the gauge charge and from the momentum $k$ in Eq.~\eqref{eq:free-matching}.
Its ultraviolet dimension $k^2$ is read from the short-distance two-point function with the opposite-charge vertex, which varies as $|X|^{-2k^2}$; see Eq.~\eqref{eq:uv-vertex}.
Appendix~\ref{sec:freecorrelatormatching} specifies the normal ordering and the subtractions needed to use these correlators for matching.

Consider $p$ mass vertices and $r$ quadratic insertions, with $p+r\ge1$, whose positions approach one another.
Each mass vertex comes from the cosine and carries charge $\sigma_s=+1$ or $-1$.
The local operator replacing this group has total charge
\begin{equation}
 k=\sum_{s=1}^{p}\sigma_s,\qquad k\equiv p\pmod2.
 \label{eq:physical-charge-count}
\end{equation}
A group can carry nonzero charge even in a neutral correlation function, because vertices outside it can supply the opposite charge.
Let $d$ be the total number of derivatives on the resulting local operator.

The power of $a$ comes from two factors.
The first is the short-distance operator product. The $p$ unit-charge vertices have total ultraviolet dimension $p$, whereas the resulting charge-$k$ operator with $d$ derivatives has dimension $k^2+d$.
Their difference gives a factor $a^{k^2+d-p}$. Quadratic insertions have zero algebraic ultraviolet dimension and contribute only logarithms to this factor.
The second factor is the integration measure. Of the $p+r$ interaction positions, one locates the resulting local operator and the other $p+r-1$ are relative positions.
Each relative position has two coordinates, giving $a^{2(p+r-1)}$ when all relative distances are rescaled by $a$.
Including the coupling factors, a possible contribution at order $m^p(g^2)^r$ therefore has the form
\begin{equation}
 h_i^{[p,r]}(a,q)\sim
 m^p(g^2)^r a^{\lambda}P_{i,p,r}(\ln(aq)),
 \qquad \lambda=k^2+d+p+2r-2.
 \label{eq:physical-power-count}
\end{equation}
Here $h_i^{[p,r]}$ denotes the contribution to $h_i$ at order $m^p(g^2)^r$, including these coupling factors; $\lambda$ is the power of $a$, and $P_{i,p,r}$ is a polynomial in $\ln(aq)$.
The scale $q$ is the fixed renormalization scale in Eq.~\eqref{eq:symanzik}.
We assume a finite polynomial $P_{i,p,r}$ at each perturbative order, after subtracting the continuum contribution and terms already reproduced by the effective action.
Appendix~\ref{sec:evenpowers} gives this counting directly in the matching equations.

Independent reflections of space and Euclidean time require an even number of each type of derivative in a bulk action term, so $d$ is even.
Since $k^2$ and $k$ have the same parity, Eq.~\eqref{eq:physical-charge-count} gives
\begin{equation}
 \lambda\equiv k+p\equiv0\pmod2.
 \label{eq:physical-even-powers}
\end{equation}
The free-propagator expansion, Eq.~\eqref{eq:free-matching}, also contains only even powers.
Insertions of already determined corrections and their subtractions preserve this parity.
After matching the continuum parameters and removing vacuum terms, the first possible positive power is therefore $a^2$ within this local expansion.

For example, two opposite-charge mass vertices have $p=2$, $r=0$ and $k=0$.
Their contribution to a two-derivative operator has $d=2$ and hence the possible dependence $m^2a^2$ times logarithms.
Appendix~\ref{sec:pairprojection} expresses the corresponding coefficients of $(\partial_x\phi)^2$ and $(\partial_\tau\phi)^2$ as subtracted integrals of free-boson correlation functions.
Three mass vertices with total charge one and no derivatives similarly allow a contribution proportional to $m^3a^2$ to the fundamental vertex coefficient.
Such counting identifies allowed powers; cancellations can remove a logarithmic term or the whole contribution.

\Needspace{8\baselineskip}
\subsection{From local corrections to the critical mass}
\label{sec:criticalshift}

The matching construction in Appendix~\ref{sec:freecorrelatormatching} determines short-distance coefficients order by order, subject to its stated assumptions.
Matching determines how the lattice modifies the continuum action. To infer the shift of the critical mass, we also need the response of the interacting theory to those modifications.
This response is nonperturbative.
At fixed $g>0$ and $q/g$, define $U(\{h_i\})$ as the critical value of $u=m/g$ in the continuum theory with the added operators.
We assume that this value has a Taylor expansion to the orders used,
\begin{equation}
 U(\{h_i\})=c_0+\sum_i A_i h_i+\sum_{i,j}B_{ij}h_i h_j+\cdots.
 \label{eq:criticalresponse}
\end{equation}
At fixed $g$ and $q$, $A_i=\left.\partial U/\partial h_i\right|_{h=0}$ and $B_{ij}=\left.\tfrac12\partial^2U/(\partial h_i\partial h_j)\right|_{h=0}$, where $h=0$ means that all $h_i$ vanish.
Only the location of the transition is assumed to be smooth in the $h_i$; the mass gap itself may be nonanalytic there.

There is a self-consistency condition: the coefficients $h_i$ depend on the mass ratio $u$, so at the lattice critical point they must be evaluated at that same critical ratio.
Thus
\begin{equation}
 u_*(a)=U\bigl(\{h_i(a,q;u_*(a))\}\bigr).
 \label{eq:criticalcondition}
\end{equation}
We infer the form of the critical-mass expansion by expanding
Eq.~\eqref{eq:criticalcondition} formally about $u_*=c_0$.
The matching coefficients start at order $a^2$, up to logarithmic
factors. At leading order, the shift $u_*-c_0$ is therefore
$\sum_i A_i h_i(a,q;c_0)$.
Nonlinear terms in the response and the dependence of the matching
coefficients on $u_*-c_0$ contribute at higher orders.
Repeated substitution preserves even powers of $a$ multiplied by
polynomials in $\ln(aq)$.
Under the stated assumptions, this gives the asymptotic form
\begin{equation}
 u_*(a)\sim c_0+\sum_{n\ge1}z^{2n}P_{2n}(\ln z),\qquad z=ga.
 \label{eq:criticaleven}
\end{equation}
Here $u_*(a)$ is the dimensionless lattice mass denoted by $m_*(a)$ in the numerical sections.
Because $q/g$ is fixed, replacing $\ln(aq)$ by $\ln z$ only changes polynomial coefficients.

Equation~\eqref{eq:criticaleven} motivates the even powers used in Sec.~\ref{sec:contmodel}.
The truncations, the omission of logarithms at order $z^4$ and beyond, the retention criterion and the weights used for the numerical result remain fitting choices, whose sensitivity is reported in Secs.~\ref{sec:logcheck} and~\ref{sec:budget}.

\section{Conclusion}
\label{sec:conclusion}

We have determined the critical mass at $\theta=\pi$ with a bosonized lattice Hamiltonian and charge-conjugation-preserving uniform MPS.
The free-chain reduced density matrix supplies a local oscillator basis, and the odd-sector transfer-matrix spectrum permits an extrapolation in bond dimension.
The transverse-field Ising chain provides a test of this extrapolation against a known critical point.

We obtain $(m/g)_c=\mcvalerr$ by averaging nineteen continuum fits with Akaike weights.
These fits use six nested truncations in even powers through $(ga)^{12}$, a free coefficient of $(ga)^2\ln(ga)$, and five lattice-spacing ranges.
The uncertainty combines the propagated input uncertainties with the spread among the fitted continuum values.
Agreement with staggered-fermion results provides a check using a different lattice formulation.

The result is conditional on the fitting functions and on the averaging prescription.
The fitting functions contain no logarithm at order $(ga)^4$ or beyond.
Fits that include such logarithms give averages within $1\times10^{-6}$ of the result, with quoted uncertainties of $8\times10^{-6}$ to $14\times10^{-6}$.
Adding logarithmic terms of order $(ga)^4$ with coefficients up to $4\times10^{-3}$ to the data moves the result by at most $5.2\times10^{-6}$.
The quoted uncertainty therefore assumes that these coefficients are not much larger than those that the data determine at lower order.
Without the logarithm at order $(ga)^2$ the average is higher by $6.4\times10^{-6}$.
The Akaike weights specify how we combine fits; they are not probabilities of the fitting functions because the input uncertainties are systematic estimates from a deterministic calculation.

The finite-$D$ extrapolation remains a potential source of sensitivity because the available data do not constrain extensions of the leading scaling law.
Additional large-$D$ data would provide a more direct test, while a calculation of the logarithmic matching coefficients would test the assumptions used in the continuum fit.

The agreement with staggered-fermion calculations supports the use of this bosonic Hamiltonian for continuum studies of the critical point.
The oscillator basis and exact interaction matrix elements specify a finite local representation whose truncation can be tested at fixed lattice spacing.
The critical masses at individual spacings provide reference values for other tensor-network calculations and, after accounting for finite-volume and local-truncation effects, for quantum simulations of the same Hamiltonian.
These comparisons would test the Hamiltonian formulation before the further extrapolation to the continuum.

\begin{samepage}
\section*{Data and code availability}
The data and code supporting this work are available on Zenodo~\cite{Okuda:2026bosonizeddata}.
The release includes the unrounded continuum-fit inputs, the numerical uncertainty components, the archived transfer-matrix data and their source records, the local-cutoff comparisons, the analysis scripts, the fit tables, the continuum plot, and the Julia simulation code with its package environment.
The analysis scripts reproduce the primary result, the tests of logarithmic terms and other fitting functions, the input-mass sensitivity checks, and the local-cutoff mass-shift estimates from the supplied inputs.
The release includes the native transverse-field Ising-chain spectrum records and scripts that reconstruct the alternative per-spacing fits, the extended scaling fit, the two-stage mass estimates, and the Ising-chain tests. Reproduction starts from saved transfer-matrix observables and does not rerun ground-state optimization.
\par\end{samepage}

\Needspace{12\baselineskip}
\begin{samepage}
\section*{Acknowledgments}
I thank Kohei Fujikura for useful discussion.
Computations ran on an Apple M2~Ultra Mac~Studio (24 cores, 192\,GB),
the Heian computing cluster (AMD EPYC~9755 nodes) at the Yukawa Institute
for Theoretical Physics, Kyoto University, and the Ohtaka computing cluster
at the Supercomputer Center of the Institute for Solid State Physics,
the University of Tokyo.
I used the Julia packages MPSKit.jl~\cite{mpskitjl} and
TensorKit.jl~\cite{tensorkitjl}.
I thank the Supercomputer Center, the Institute for Solid State
Physics, the University of Tokyo for the use of the facilities
(ISSPkyodo-SC-2026-A-0023).
This work was supported in part by JST PRESTO Grant Number JPMJPR23F3.

\end{samepage}

\Needspace{8\baselineskip}
\begin{samepage}
\section*{Use of generative AI}
The author used Codex (OpenAI) and Claude (Anthropic) to assist with drafting and revising the manuscript, developing and checking mathematical arguments, writing and checking simulation code and scripts for continuum extrapolation, and preparing and submitting computational jobs.
The author is responsible for the scientific claims, numerical results, and final text.
\par\end{samepage}

\clearpage
\appendix
\numberwithin{equation}{section}
\section{Free lattice theory and the local oscillator basis}
\label{sec:freeappendix}

This appendix collects the free-theory results used to normalize the
mass interaction and choose the local basis. We set the fermion mass
to $m=0$ while keeping $g>0$, so the boson remains massive, with
$\mu=g/\sqrt{\pi}$. All vacuum expectation values below refer to this
free theory on the infinite lattice.

\subsection{Normal modes on the infinite lattice}
\label{sec:freemodes}

Let $j\in\mathbb Z$ label the sites of the infinite chain, with
$[\phi_j,p_l]=i\delta_{jl}$.  The continuum momentum density is
discretized as $p_j=a\Pi_\phi(ja)$.  For continuous momenta in the
Brillouin zone $\mathcal B=(-\pi/a,\pi/a]$, introduce modes $c(k)$ with
\begin{equation*}
[c(k),c^\dagger(q)]=2\pi\delta(k-q),
\qquad [c(k),c(q)]=0.
\end{equation*}
The field expansions are
\begin{align}
\phi_j&=\int_{\mathcal B}\frac{dk}{2\pi}
\frac{c(k)e^{ikja}+c^\dagger(k)e^{-ikja}}{\sqrt{2\Omega(k)}},
\label{eq:freephi}\\
p_j&=-ia\int_{\mathcal B}\frac{dk}{2\pi}\sqrt{\frac{\Omega(k)}{2}}
\bigl(c(k)e^{ikja}-c^\dagger(k)e^{-ikja}\bigr).
\label{eq:freep}
\end{align}
Substituting into Eq.~\eqref{eq:Hlat} at $m=0$ and normal ordering in the vacuum
annihilated by all $c(k)$ gives
\begin{equation}
:\!H_0\!:_{\infty}=\int_{\mathcal B}\frac{dk}{2\pi}
\Omega(k)c^\dagger(k)c(k),
\qquad
\Omega(k)^2=\mu^2+\frac{4}{a^2}\sin^2\frac{ka}{2},
\label{eq:freeH}
\end{equation}
which yields Eq.~\eqref{eq:dispersion}.  Normal ordering removes the
vacuum energy; its finite value per site is given below in
Eq.~\eqref{eq:freeenergy}.  The modes $c(k)$ diagonalize the full
coupled chain.  They differ from the local modes $b_j$ used to truncate
the Hilbert space at each site.

Keeping the gradient term separate, the quadratic on-site Hamiltonian
is $p_j^2/(2a)+a\mu^2\phi_j^2/2$: it is an oscillator of mass $a$
and frequency $\mu$.  Thus $\mu$ is exact at every $a$; it is also the
zero-momentum frequency in Eq.~\eqref{eq:freeH}.

\subsection{Vacuum two-point functions and normal ordering}
\label{sec:freecovariances}

The subscript $0$ denotes expectation in this free infinite-chain vacuum.
Contracting the modes in Eqs.~\eqref{eq:freephi} and~\eqref{eq:freep} gives
\begin{equation}
\sigma_\phi^2=\int_{-\pi/a}^{\pi/a}\frac{dk}{2\pi}\frac{1}{2\Omega(k)},
\qquad
\sigma_p^2=a^2\int_{-\pi/a}^{\pi/a}\frac{dk}{2\pi}\frac{\Omega(k)}{2},
\qquad
\tfrac12\langle\{\phi_j,p_j\}\rangle_0=0.
\label{eq:freecovintegrals}
\end{equation}
The last equality states that the symmetrized field-momentum
correlator vanishes. The free vacuum can be chosen real in the field
basis, while $p_j$ changes sign under complex conjugation.
The ordered correlators are nonzero, as required by $[\phi_j,p_j]=i$.
The ground-state energy per site is
\begin{equation}
E_{\rm per\,site}=a\int_{-\pi/a}^{\pi/a}\frac{dk}{2\pi}\frac{\Omega(k)}{2}
=\frac{\sigma_p^2}{a}.
\label{eq:freeenergy}
\end{equation}
To evaluate these integrals, use $u=ka/2$ and the definitions
\begin{equation}
K(t)=\int_0^{\pi/2}\frac{du}{\sqrt{1-t\sin^2u}},
\qquad
E(t)=\int_0^{\pi/2}du\,\sqrt{1-t\sin^2u}.
\label{eq:ellipticdefinitions}
\end{equation}
Their argument $t$ is the elliptic parameter, rather than its square
root, the elliptic modulus.  With $t=\tilde m=(1+\varepsilon)^{-1}$,
the resulting expressions are Eq.~\eqref{eq:variances}.

Correlation functions in the Gaussian theory can be evaluated by applying Wick's theorem.
For the exponential at one site in the free vacuum, this gives, for real $\lambda$,
\begin{equation}
\langle e^{i\lambda\phi_j}\rangle_0
=e^{-\lambda^2\sigma_\phi^2/2},
\qquad
:\!e^{i\lambda\phi_j}\!:_{\infty}
=e^{\lambda^2\sigma_\phi^2/2}e^{i\lambda\phi_j}.
\label{eq:freewick}
\end{equation}
Here normal ordering means placing creation modes $c^\dagger(k)$ to
the left of annihilation modes $c(k)$.  At $\lambda=2\sqrt{\pi}$,
$\ln\mathcal N(ga)=2\pi\sigma_\phi^2$, proving
Eq.~\eqref{eq:normal-ordering}.
This prescription uses the correlated vacuum of the full chain;
normal ordering in the vacuum of a local oscillator would instead
use that oscillator's variance.

\subsection{The single-site reduced density matrix}
\label{sec:freebasis}

For any positive scale $\ell$, define the local oscillator operators
\begin{equation}
b_j=\frac{\phi_j}{2\ell}+i\ell p_j,
\qquad
\phi_j=\ell(b_j+b_j^\dagger),
\qquad
p_j=\frac{b_j-b_j^\dagger}{2i\ell},
\qquad [b_j,b_l^\dagger]=\delta_{jl}.
\label{eq:localmode}
\end{equation}
The scale $\ell$ is dimensionless and fixes the width of these basis
functions in the field coordinate; it is not a distance along the
chain.

Let $\rho_1$ be the density matrix obtained by tracing the free vacuum
over all sites except $j$. In the $\phi_j$ eigenbasis, integrating
the other field coordinates out of the ground-state wave functional
gives~\cite{Srednicki:1993im,Peschel:2009rdm}
\begin{equation}
\langle x|\rho_1|y\rangle
=\frac{1}{\sqrt{2\pi\sigma_\phi^2}}
\exp\left[-\frac{(x+y)^2}{8\sigma_\phi^2}
-\frac{\sigma_p^2}{2}(x-y)^2\right].
\label{eq:freekernel}
\end{equation}
Here $x$ and $y$ are values of the field $\phi_j$, not spatial
positions. The vanishing mixed correlator in
Eq.~\eqref{eq:freecovintegrals} has been used.
This formula specifies the entire one-site density matrix in terms
of the two variances.

The eigenfunctions of this integral kernel are harmonic-oscillator
wave functions. Their width is fixed by
$\ell^2=\ell_0^2=\sigma_\phi/(2\sigma_p)$, and the eigenvalues are
\begin{equation}
\rho_1=(1-q)\sum_{n=0}^{\infty}q^n\ket n\bra n,
\qquad
q=\frac{\sigma_\phi\sigma_p-\tfrac12}
        {\sigma_\phi\sigma_p+\tfrac12}.
\label{eq:freerdm}
\end{equation}
The diagonalization is the harmonic-oscillator calculation in
Ref.~\cite{Srednicki:1993im}, Eqs.~(3)--(5); see also
Ref.~\cite{Peschel:2009rdm}, Sec.~3.3.
The condition on $\ell_0$ is equivalently
\begin{equation}
\langle b_j^2\rangle_0=0,
\qquad
\langle b_j^\dagger b_j\rangle_0
=\sigma_\phi\sigma_p-\tfrac12=\bar n.
\label{eq:matchedmoments}
\end{equation}
Equations~\eqref{eq:freerdm} and~\eqref{eq:matchedmoments} give
Eq.~\eqref{eq:thermal}; $\ell_0^2=1/(2a\omega_0)$ gives
Eq.~\eqref{eq:omega0}.
For the coupled chain, $\sigma_\phi\sigma_p>1/2$, so $0<q<1$ and
$\rho_1$ is mixed. Its thermal-oscillator form describes entanglement
with the other sites, while the full chain remains in a pure
zero-temperature state.

The largest $n_{\max}+1$ eigenvalues correspond to $0\le n\le n_{\max}$.
Retaining these eigenvectors maximizes $\operatorname{Tr}(P\rho_1)$
over projectors $P$ of that rank, and discards probability
$q^{n_{\max}+1}$. This proves optimality for the free one-site density
matrix.

For $ga\ll1$, the elliptic integrals give
\begin{equation}
\sigma_\phi^2\simeq\frac{1}{2\pi}\ln\frac{8\sqrt\pi}{ga},
\qquad \sigma_p^2\simeq\frac{2}{\pi}.
\label{eq:freeasymptotics}
\end{equation}
These limits yield the asymptotic frequency in
Eq.~\eqref{eq:omega0}. The variances receive contributions from the
whole Brillouin zone, which explains why the basis frequency differs
from the zero-momentum mass $\mu$.

\section{The \texorpdfstring{$\delta$}{delta} extrapolation in the transverse-field Ising chain}
\label{sec:tfim}

To test the $\delta$ extrapolation where the answer is known, we applied the method of Sec.~\ref{sec:delta} to the transverse-field Ising chain $H=-\sum_jZ_jZ_{j+1}-h\sum_jX_j$, where $X_j$ and $Z_j$ are Pauli matrices at site $j$ and $h$ is the transverse field. 
It belongs to the same universality class. 
For $h>1$ the exact inverse correlation length of $Z_j$ is $\ln h$~\cite{Pfeuty:1970qrn,Barouch:1971ywx}. 
We computed uniform MPS with the $\mathbb Z_2$ symmetry $\prod_jX_j$ imposed exactly, at twenty-one couplings $0.9999\le h\le1.03$ and bond dimensions $8\le D\le256$, and took $\epsilon_1$ and $\delta$ from the odd sector.
The states were grown and analyzed as for the bosonized chain.
The data contain $383$ computed states at $329$ distinct $(h,D)$ pairs: $54$ pairs were computed in two separate runs with different sequences of bond dimensions.
The interpolation and cubic fits below retain both results at each repeated pair, with equal weight for each computed state.

At $h=1$, $\epsilon_1/\delta=0.278$--$0.281$ for $32\le D\le256$, close to the values of $c_1$ found in Sec.~\ref{sec:delta}. 
Let $\epsilon^{\rm even}_1$ be the smallest positive value of $-\ln|\lambda|$ in the even block, after excluding its dominant eigenvalue $\lambda=1$.
The four lowest levels of the transfer matrix, $\{0,\epsilon_1,\epsilon_2,\epsilon^{\rm even}_1\}$, have ratios approximately $\{0,\,0.177,\,0.812,\,1\}$ over the measured bond dimensions, consistent with $\{0,\,0.1769,\,0.8130,\,1\}$ of Ref.~\cite{Schneider:2024ipg}.
Here we use $t=\ln h$ as the signed scaling parameter: it agrees with the exact inverse correlation length for $h>1$ and changes sign at $h=1$. With $x=t/\delta$, the points with $D\ge32$ and $x\ge-0.21$ fall on one curve,
\begin{equation}
F(x)=x+0.332-\frac{0.0532}{1+0.949\,x},
\label{eq:tfimF} 
\end{equation}
with an rms deviation of $0.0011$ for $-0.21\le x\le10.3$.
Equation~\eqref{eq:tfimF} is a least-squares fit, with equal weights, of the form $F(x)=x+B-(B-F_0)/(1+bx)$ to the $272$ computed states at $242$ distinct $(h,D)$ pairs with $D\ge32$ and $x\ge-0.21$, with $B$, $F_0$ and $b$ free. 
The form has $F(0)=F_0$ and approaches $x+B$ at large $x$; it interpolates between the two limits
and is not derived. 
Expanding \(F(x)\) about \(x=0\) gives \(F_0\simeq0.279\) and \(F_1=F'(0)\simeq1.05\). At fixed \(h>1\), retaining these terms gives
\[
\epsilon_1\simeq F_0\delta+F_1\ln h.
\]
This motivates fitting \(\epsilon_1\) against \(\delta\) by a straight line at fixed \(h\), using data with \(x=(\ln h)/\delta\ll1\). Within this approximation, the fitted line has a \(\delta=0\) intercept \(F_1\ln h\simeq1.05\ln h\), about \(5\%\) above the exact inverse correlation length.
The joint fit of Eq.~\eqref{eq:jointfit}, with $h$ in place of $m$ and applied to couplings within $3\times10^{-4}$ of $h=1$ with $D\ge32$ and $x\ge-0.2$, locates the critical point within $7\times10^{-7}$ of $h=1$. 
We also use the Ising data to test the nonlinear coefficients in Eq.~\eqref{eq:extfit}.
First, we fit a cubic polynomial in $x$ to the $122$ computed states at $116$ distinct $(h,D)$ pairs with $D\ge128$ and $-0.2\le x\le1.1$; the rms deviation is $0.0008$.
We then use this polynomial to generate synthetic data at the masses and bond dimensions used for the bosonized chain, and fit those data with Eq.~\eqref{eq:extfit}.
The resulting coefficients are $\phi_2=-0.028(3)$ and $\phi_3=0.008(2)$, compared with $-0.025(2)$ and $0.006(2)$ for the bosonized chain. The Ising uncertainties are the standard deviations from 80 bootstrap resamples of the selected states; they describe the scatter of the interpolation, not a statistical uncertainty of the deterministic MPS calculation. This comparison is specific to the cubic interpolation and is not a universal prediction of the Ising scaling function.
If only the quadratic correction is fitted, the values of $\phi_2$ are $-0.0170(7)$ and $-0.0181(8)$, respectively.

\FloatBarrier
\clearpage
\section{Matching coefficients from free-boson correlation functions}
\label{sec:freecorrelatormatching}

This appendix explains how to determine the effective-action coefficients $h_i$ introduced in Sec.~\ref{sec:c2}.
The matching procedure compares lattice and continuum correlation functions in three steps: put them in the same normal-ordering convention, subtract the effects of coefficients already determined, and express the remaining short-distance difference as local operator insertions.
Section~\ref{sec:matchingconventions} sets up the correlation functions and subtractions.
Section~\ref{sec:localprojection} defines how to extract a local contribution and solve for its coefficient.
Section~\ref{sec:pairprojection} works through a pair of mass vertices, and Sec.~\ref{sec:evenpowers} derives the power count in Eq.~\eqref{eq:physical-power-count}.

The matching below keeps the original scalar field and its normalization.
Total derivatives can be removed, but operators are not eliminated by field redefinitions or by using the equations of motion.
Such eliminations would also change the field insertions whose correlation functions are being matched.
The power counting below does not require a complete list of independent local operators.

\subsection{Free-boson correlators and normal ordering}
\label{sec:matchingconventions}

We use the Gaussian reference action
\begin{equation}
 S_{\rm G}=\frac12\int d^2X\,(\partial\phi)^2,\qquad \phi\in\mathbb R,
 \label{eq:massless-reference}
\end{equation}
and treat both the cosine interaction and the quadratic potential as insertions.
Here $(\partial\phi)^2=(\partial_x\phi)^2+(\partial_\tau\phi)^2$, and $\langle\cdots\rangle_{\rm G}$ denotes expectation with the Gaussian action $S_{\rm G}$.
Two free theories appear in this paper. The physical theory at zero fermion mass has $m=0$ and $g>0$, so its boson has mass $\mu$.
The Gaussian reference theory used here has $m=g=0$. We use its correlation functions to expand both interactions perturbatively.
Set $\beta=2\sqrt\pi$.
We denote normal ordering with respect to the massless free-boson propagator by $:\cdots:_q$:
\[
 H_q(X)=-\frac1{2\pi}\ln(q|X|),\qquad X\ne0.
\]
The scale $q$ fixes the additive constant in $H_q$.
An infrared regulator for correlation functions is introduced below.
The normalized vertex operators are
\begin{equation}
 \widehat V_{k,q}=q^{k^2}:e^{ik\beta\phi}:_q,
 \qquad
 \langle\widehat V_{k,q}(X)\widehat V_{-k,q}(0)\rangle_{\rm G}
 =|X|^{-2k^2}.
 \label{eq:uv-vertex}
\end{equation}
The second equality holds after removal of the infrared regulator at separated points.
To compare with Eq.~\eqref{eq:Hcont}, retain its notation $:\cdots:_\infty$ for normal ordering in the infinite-volume free vacuum of mass $\mu=g/\sqrt\pi$.
The continuum covariance has the short-distance form
\[
 G_\mu(X)=\frac1{2\pi}K_0(\mu|X|)
 =-\frac1{2\pi}\ln\!\left(\frac{e^\gamma\mu}{2}|X|\right)+o(1),
 \qquad |X|\to0,
\]
where $K_0$ is the modified Bessel function of the second kind.
Using the same short-distance regulator in both prescriptions gives
\[
 \begin{aligned}
 :e^{ik\beta\phi}:_q
 &=\left(\frac{e^\gamma\mu}{2q}\right)^{k^2}
       :e^{ik\beta\phi}:_\infty,\\
 \widehat V_{k,q}
 &=\left(\frac{e^\gamma\mu}{2}\right)^{k^2}
       :e^{ik\beta\phi}:_\infty.
 \end{aligned}
\]
These continuum relations hold for any $q$.
The normal-ordered exponentials themselves agree at $q=e^\gamma\mu/2$; at $q=\mu$ their conversion factor is $(e^\gamma/2)^{k^2}$.
No such special choice of $q$ is required for matching.
Writing
\begin{equation}
 \mathcal V_{\theta,q}=\frac12\sum_{\sigma=\pm1}e^{-i\sigma\theta}\widehat V_{\sigma,q},
 \qquad Q_q=:\phi^2:_q,
 \qquad
 S_{\rm pert}=-\frac m\pi\int d^2X\,\mathcal V_{\theta,q}
              +\frac{g^2}{2\pi}\int d^2X\,Q_q
 \label{eq:massless-perturbations}
\end{equation}
reproduces the continuum interactions in Eq.~\eqref{eq:Hcont}, up to a vacuum constant.
Indeed, the vertex conversion gives the cosine coefficient in that equation, while
\[
 Q_q=:\phi^2:_\infty+\frac1{2\pi}\ln\!\left(\frac{2q}{e^\gamma\mu}\right)
\]
differs from the massive normal-ordered quadratic field only by a constant.
The integer charges arise from expanding the cosine interaction; they do not mean that the reference field is compact.

The massless field and its undifferentiated powers require an infrared prescription.
We introduce the same temporary mass $\epsilon>0$ in the two Gaussian calculations, with momentum-space covariances
\begin{equation}
 \widetilde G_{0,\epsilon}(p_x,p_\tau)
 =\frac1{p_\tau^2+p_x^2+\epsilon^2},\qquad
 \widetilde G_{a,\epsilon}(p_x,p_\tau)
 =\frac1{p_\tau^2+4a^{-2}\sin^2(ap_x/2)+\epsilon^2}.
 \label{eq:matching-covariances}
\end{equation}
Here $\epsilon$ regulates the zero mode and is not the physical mass $\mu$.
The variables $p_x$ and $p_\tau$ are spatial momentum and Euclidean frequency, respectively; a tilde denotes a Fourier transform.
The covariance $G_{j,\epsilon}$ is the free two-point function, with $j=0$ for the continuum and $j=a$ for the spatial lattice.
We denote the corresponding quadratic action by $S_{{\rm G},j,\epsilon}$.
Let $G_{a,\epsilon}(X)$, with $X=(na,\tau)$, denote the inverse Fourier transform of $\widetilde G_{a,\epsilon}$, with $p_x\in[-\pi/a,\pi/a]$ and $p_\tau\in\mathbb R$.
The massless lattice propagator with the same additive-constant convention as $H_q$ is
\[
 H_{a,q}(X)=\lim_{\epsilon\to0}\left[
 G_{a,\epsilon}(X)+\frac1{2\pi}\ln\frac{e^\gamma\epsilon}{2q}
 \right].
\]
At fixed nonzero separation, $H_{a,q}(X)$ tends to $H_q(X)$ as $a\to0$.
The free-chain covariance in Appendix~\ref{sec:freecovariances}, with $\mu$ replaced by $\epsilon$, gives
\[
 G_{a,\epsilon}(0)=\frac1{2\pi}\ln\frac8{a\epsilon}+o(1),
 \qquad
 H_{a,q}(0)=\frac1{2\pi}\ln\frac{4e^\gamma}{aq},
 \qquad \epsilon\to0.
\]
Normal ordering on this spatial lattice is therefore defined by
\[
 \begin{aligned}
 :e^{ik\beta\phi(X)}:_{q,a}
 &=e^{2\pi k^2H_{a,q}(0)}e^{ik\beta\phi(X)},\\
 Q_{q,a}(X)&=:\phi(X)^2:_{q,a}=\phi(X)^2-H_{a,q}(0),\\
 \widehat V_{k,q,a}&=q^{k^2}:e^{ik\beta\phi}:_{q,a},
 \qquad
 \mathcal V_{\theta,q,a}=\frac12\sum_{\sigma=\pm1}e^{-i\sigma\theta}\widehat V_{\sigma,q,a}.
 \end{aligned}
\]
To distinguish the physical lattice prescription from its continuum counterpart, write the former as $:\cdots:_{\infty,a}$ in this subsection.
Equation~\eqref{eq:normal-ordering} then gives
\[
 :e^{ik\beta\phi}:_{\infty,a}
 =\exp\!\left\{2\pi k^2\bigl[G_{a,\mu}(0)-H_{a,q}(0)\bigr]\right\}
 :e^{ik\beta\phi}:_{q,a},
 \qquad e^{2\pi G_{a,\mu}(0)}=\mathcal N(ga).
\]
Consequently, the interaction obtained from Eq.~\eqref{eq:Hlat}, up to a vacuum constant, is
\[
 S_{{\rm pert},a}=-\frac m\pi Z_a\int_a d^2X\,\mathcal V_{\theta,q,a}
                 +\frac{g^2}{2\pi}\int_a d^2X\,Q_{q,a},
 \qquad
 Z_a=\frac{a\mu}{8}\mathcal N(ga).
\]
Here $\int_a d^2X=a\sum_n\int d\tau$.
The factor $Z_a$ tends to one as $a\to0$, but its spacing-dependent corrections must be retained.
For uniform notation below, set $Z_0=1$, $\mathcal V_{\theta,q,0}=\mathcal V_{\theta,q}$ and $Q_{q,0}=Q_q$ in the continuum.
The scales $q$ and $\epsilon$ serve different purposes. The normal-ordering scale $q$ fixes the operator subtractions and is held fixed.
The temporary mass $\epsilon$ regulates the infrared behavior of the correlation functions and is removed at the end.
We assume that the local coefficients have a finite limit as $\epsilon\to0$ after the lattice--continuum difference and all known effective-action contributions have been subtracted.
The individual integrated correlators need not have finite limits on their own.
The hierarchy $|m|,g\ll q\ll a^{-1}$ separates physical scales from the short distances used for matching; it does not assume that the critical state is weakly coupled.
At the coarsest fitted spacings, where $a^{-1}$ is only slightly larger than $g$, this hierarchy is marginal.

For matching, use elementary fields at distinct positions $y_{A,b}$:
\[
 \mathcal A_A=\prod_{b=1}^{n_A}\phi(y_{A,b}).
\]
The label $A$ specifies the correlation function through its number of fields $n_A$ and their positions $y_{A,b}$; $b$ labels each insertion.
We call these fields external because their positions are held fixed while the interaction positions $X_s$ and $Y_t$ are integrated.
For example, $n_A=2$ gives the two-point function of $\phi(y_{A,1})$ and $\phi(y_{A,2})$.
We match the correlation functions with a source coupled linearly to this same elementary field in the lattice and continuum theories.
We keep corrections to the coefficients of the quadratic derivative terms as effective-action terms rather than absorbing them into a rescaling of $\phi$.
A subsequent field redefinition would require transforming these insertions as well.
The subscript $c$ denotes the connected part of a correlation function, obtained by subtracting products of correlators for disconnected groups of insertions.
Each external field $\phi(y_{A,b})$, each mass vertex, and each quadratic insertion is treated as a separate insertion.
In particular, $\mathcal A_A$ abbreviates a product of external fields; it is not treated as one composite insertion when taking the connected part.
In the Gaussian calculation, the connected part keeps the Wick-contraction patterns that link all these separate insertions.
The Gaussian integral with $p$ mass vertices and $r$ quadratic insertions, before including the factor $Z_j^p$, is
\begin{equation}
 \begin{split}
 \mathcal I_{j,A}^{[p,r]}
 ={}&\frac{(m/\pi)^p[-g^2/(2\pi)]^r}{p!\,r!}
 \int_j\prod_{s=1}^{p}d^2X_s\prod_{t=1}^{r}d^2Y_t\\
 &\times\left\langle\mathcal A_A
       \prod_{s=1}^{p}\mathcal V_{\theta,q,j}(X_s)
       \prod_{t=1}^{r}Q_{q,j}(Y_t)
       \right\rangle_{G_{j,\epsilon},c},\qquad j=a,0.
 \end{split}
 \label{eq:massless-correlator-expansion}
\end{equation}
The measures are $\int_a d^2X=a\sum_n\int d\tau$ and $\int_0d^2X=\int dx\,d\tau$.
Every expectation value on the right is taken in the regulated Gaussian theory and is evaluated by expanding the composite insertions in elementary fields and applying Wick's theorem.
The signs follow from $e^{-S_{\rm pert}}$; disconnected vacuum factors are absent from the connected part.
On the lattice, each mass vertex also carries the normal-ordering factor $Z_a$.
Because this factor depends on $g^2$, fewer than $r$ explicit quadratic insertions can still contribute at order $m^p(g^2)^r$.
To collect all contributions at that order, expand $Z_j^p$ as well:
\[
 \mathcal C_{j,A}^{[p,r]}
 =\left[\sum_{s=0}^{r}Z_j^p\mathcal I_{j,A}^{[p,s]}\right]^{[p,r]}.
\]
The brackets $[\cdots]^{[p,r]}$ retain the term proportional to $m^p(g^2)^r$, including these factors and any logarithms.
Thus $\mathcal C_{0,A}^{[p,r]}=\mathcal I_{0,A}^{[p,r]}$, whereas the lattice coefficient can also receive contributions with fewer than $r$ quadratic insertions through the expansion of $Z_a^p$.
Replacing $Z_a$ by one would discard these contributions from normal ordering.
This is a formal expansion in $m$ and $g^2$, not an assertion of convergence at the critical mass.

Let $\langle\mathcal A_A\rangle_{j,\epsilon,c}$ denote the connected correlation function in the interacting theory with action $S_{{\rm G},j,\epsilon}+S_{{\rm pert},j}$, using the same infrared regulator and normal-ordering convention as above.
Equation~\eqref{eq:massless-correlator-expansion}, with the factors $Z_j^p$ included as above, gives its formal expansion as
\[
 \langle\mathcal A_A\rangle_{j,\epsilon,c}
 =\sum_{p,r\ge0}\mathcal C_{j,A}^{[p,r]},
 \qquad
 \mathcal C_{j,A}^{[0,0]}
 =\langle\mathcal A_A\rangle_{G_{j,\epsilon},c}.
\]
The term $p=r=0$ is the free connected correlation function; the terms with $p+r\ge1$ give the interaction corrections at fixed $\epsilon$.

Matching is recursive: before determining a new coefficient, we subtract every contribution already generated by known coefficients.
Fix the power of $a$ to be matched, and write $h_i^{\rm known}$ for the terms determined at lower powers of $a$ or, at the same power, at lower total order in $m$ and $g^2$.
Their contribution to the continuum connected correlation function is
\[
 \begin{aligned}
 \mathcal C_{A,\rm known}^{[p,r]}
 =\Bigg[{}&-\sum_i h_i^{\rm known}\int d^2X\,
       \langle\mathcal A_A\mathcal O_i^R(X;q)\rangle_{0,\epsilon,c}\\
 &+\frac12\sum_{i,k}h_i^{\rm known}h_k^{\rm known}
       \int d^2X\,d^2Y\,
       \langle\mathcal A_A\mathcal O_i^R(X;q)
                    \mathcal O_k^R(Y;q)\rangle_{0,\epsilon,c}
       +\cdots\Bigg]^{[p,r]}.
 \end{aligned}
\]
The brackets use the same extraction of the $m^p(g^2)^r$ term as above, now also restricted to the chosen power of $a$; logarithms of $aq$ are retained.
The expectation values here are in the interacting continuum theory.
Expanding its mass and quadratic interactions as in Eq.~\eqref{eq:massless-correlator-expansion} expresses them in Gaussian correlation functions.
Every external operator and every $\mathcal O_i^R$ insertion is a separate entry when taking the connected part, and products of insertions use the same subtraction prescription at $q$.
The signs and the factor $1/2$ follow from expanding $e^{-\sum_i h_i^{\rm known}\int\mathcal O_i^R}$.
In particular, two known order-$a^2$ coefficients in the second term produce an order-$a^4$ contribution that must be subtracted before determining a new order-$a^4$ coefficient.

\subsection{Local projection and matching equations}
\label{sec:localprojection}

The local projection replaces a group of nearby interaction insertions by an operator at a single point.
It does this by Taylor-expanding the group's connections to distant insertions while retaining the contractions within the group in weighted integrals.
The operation must be performed before integrating out the relative coordinates. The following definitions make this procedure precise.
For a chosen group of $N\ge2$ interaction points at $Z_1,\ldots,Z_N$, expand about their average position $X=N^{-1}\sum_{s=1}^N Z_s$.
Choose $N-1$ independent relative two-vectors, collected into $\xi\in\mathbb R^{2(N-1)}$.
For a pair, write $X_\pm=X\pm r/2$ and $\xi=r$.
In a region where this group stays apart from the external insertions and any other interaction points, a Wick-contraction term can be written as
\begin{equation}
 \bigl\langle T_k^R(\xi),F_A(X,\xi)\bigr\rangle_\xi.
 \label{eq:cluster-contraction-pairing}
\end{equation}
Here $k$ is the sum of the vertex charges in the chosen group.
For $p$ mass vertices from Eq.~\eqref{eq:massless-correlator-expansion}, $k=\sum_{s=1}^p\sigma_s$ with $\sigma_s=\pm1$; quadratic insertions carry zero charge.
For each charge assignment and Wick-contraction term, $T_k^R$ contains the contractions within the nearby group.
The factor $F_A$ describes its connections to the distant insertions and is smooth in the relative coordinates.
We fix the total charge $k$ and select a derivative order $d$ before applying the projection.
The brackets in Eq.~\eqref{eq:cluster-contraction-pairing} mean that the distribution $T_k^R$ acts on the function $F_A$; they do not denote another expectation value.
For a smooth $F$ of compact support, write $\langle T,F\rangle_\xi:=T[F]$.
If $T$ is a locally integrable function, this means
\begin{equation}
 \langle T,F\rangle_\xi
 =\int d^{D_\xi}\xi\,T(\xi)F(\xi),
 \qquad D_\xi=2(N-1).
 \label{eq:distribution-pairing}
\end{equation}
For a lattice pair, the corresponding definition is
\begin{equation}
 \langle T,F\rangle_r
 =a\sum_{n\in\mathbb Z}\int dr_\tau\,
       T(na,r_\tau)F(na,r_\tau).
 \label{eq:lattice-distribution-pairing}
\end{equation}
For distributions concentrated at coincidence, the same notation gives
\begin{equation}
 \langle\delta^{(D_\xi)},F\rangle_\xi=F(0),
 \qquad
 \langle\partial_\xi^\alpha\delta^{(D_\xi)},F\rangle_\xi
 =(-1)^{|\alpha|}\partial_\xi^\alpha F(0),
 \label{eq:delta-derivative-pairing}
\end{equation}
where $\alpha$ is a multi-index and $|\alpha|$ is its total derivative order.
Explicitly, for $\alpha=(\alpha_1,\ldots,\alpha_{D_\xi})$ with nonnegative integer entries,
\begin{equation}
 |\alpha|=\sum_{b=1}^{D_\xi}\alpha_b,\qquad
 \alpha!=\prod_{b=1}^{D_\xi}\alpha_b!,\qquad
 \xi^\alpha=\prod_{b=1}^{D_\xi}\xi_b^{\alpha_b},\qquad
 \partial_\xi^\alpha=\prod_{b=1}^{D_\xi}\partial_{\xi_b}^{\alpha_b}.
 \label{eq:multi-index-notation}
\end{equation}
The distribution $\delta^{(D_\xi)}$ is the Dirac delta function in $D_\xi$ dimensions.
The contractions can be singular when interaction points coincide.
The superscript $R$ means that they have been defined as distributions at those coincidences using the fixed subtraction prescription at $q$.
If a smaller group of points inside the selected group also becomes coincident, its contribution is subtracted first and retained among the known terms in $\mathcal C_{A,\rm known}$.
Normal ordering fixes the contractions within a single operator; the finite contact terms in products of different operators are fixed by the operator and product conventions used on both sides of the matching.
Contact terms are distributions supported where insertion points coincide, such as the delta functions and their derivatives above.

Fix a smooth, rotation-invariant function $\chi(\eta)$ of the dimensionless relative coordinates $\eta\in\mathbb R^{D_\xi}$, equal to one for $|\eta|\le1$ and zero for $|\eta|\ge2$.
Its choice, the relative-coordinate convention and $q$ are held fixed as $a$ changes.
The function $\chi(q\xi)$ restricts the Taylor expansion to relative distances of order $q^{-1}$.
To select operators with $d$ derivatives, retain the Taylor terms of total degree $d$:
\begin{equation}
 (J_{\chi,q}^{(d)}F)(\xi)
 =\chi(q\xi)\sum_{|\alpha|=d}
       \frac{\xi^\alpha}{\alpha!}\,\partial_\xi^\alpha F(0),
 \qquad
 \bigl\langle\mathcal P_{\rm loc}^{(k,d)}T_k^R,F\bigr\rangle
 :=\bigl\langle T_k^R,J_{\chi,q}^{(d)}F\bigr\rangle.
 \label{eq:local-projection-definition}
\end{equation}
Equivalently, using the action of delta derivatives in Eq.~\eqref{eq:delta-derivative-pairing},
\begin{equation}
 \mathcal P_{\rm loc}^{(k,d)}T_k^R
 =(-1)^d\sum_{|\alpha|=d}
 \frac{\langle T_k^R,\chi(q\xi)\xi^\alpha\rangle}{\alpha!}
       \partial_\xi^\alpha\delta^{(D_\xi)}(\xi).
 \label{eq:local-projection-distribution}
\end{equation}
Equation~\eqref{eq:local-projection-distribution} shows the result explicitly: the group is replaced by derivatives of a delta function, with coefficients given by moments of its internal contractions.
Only the smooth external factor has been Taylor-expanded.
Since $\chi=1$ near coincidence, applying the operation again leaves these local terms unchanged; hence $(\mathcal P_{\rm loc}^{(k,d)})^2=\mathcal P_{\rm loc}^{(k,d)}$.
For the insertions in Eq.~\eqref{eq:massless-correlator-expansion}, Taylor degree $d$ gives $d$ derivatives on the resulting local fields.
If a known insertion already leaves $d_0$ derivatives on uncontracted fields, use Taylor degree $d-d_0$ to select total derivative order $d$.
On an already local insertion the projection simply retains its charge and derivative sector.

The projection $\mathcal P_{\rm loc}^{(k,d)}$ acts on the relative-coordinate distribution $T_k^R$ at this fixed charge.
To define an operation on the integral representation of a correlation function, fix the sector $(k,d)$ and set
\begin{equation}
 \begin{aligned}
 \mathcal P_{\rm loc}\!\left[
   \int d^2X\,\langle T_k^R,F_A(X,\cdot)\rangle_\xi\right]
 &:=\int d^2X\,
       \langle\mathcal P_{\rm loc}^{(k,d)}T_k^R,F_A(X,\cdot)\rangle_\xi\\
 &=\int d^2X\,
       \langle T_k^R,J_{\chi,q}^{(d)}F_A(X,\cdot)\rangle_\xi.
 \end{aligned}
 \label{eq:local-projection-on-integrals}
\end{equation}
Here $F_A(X,\cdot)$ means the function of $\xi$ with $X$ held fixed.
For a sum of Wick-contraction terms, apply the right-hand side to each term and add the results; a lattice term includes the corresponding sum over the common spatial coordinate.
Thus $\mathcal P_{\rm loc}$ acts inside the integrals, on their relative-coordinate dependence.
It cannot be reconstructed from the final value of an already integrated correlator: the integral representation must be kept until the projection is complete.
We use the same charge and derivative sector for the residual and for every operator response in Eq.~\eqref{eq:free-correlator-matching}.

The two pairs of labels answer different questions.
The perturbative label $[p,r]$ specifies the order $m^p(g^2)^r$ of a contribution.
The projection label $(k,d)$ specifies the charge and derivative count of the local operator it produces.
For example, two opposite-charge mass vertices contributing to a two-derivative operator have $[p,r]=[2,0]$ and $(k,d)=(0,2)$; Sec.~\ref{sec:pairprojection} works through this case.

The residual to be matched is
\begin{equation}
 \mathcal D_A^{[p,r]}
 =\mathcal P_{\rm loc}\!\left[
      \mathcal C_{a,A}^{[p,r]}-\mathcal C_{0,A}^{[p,r]}
      -\mathcal C_{A,\rm known}^{[p,r]}
    \right].
 \label{eq:matching-residual}
\end{equation}
All terms on the right are restricted to the power of $a$ being matched.
The projection in Eq.~\eqref{eq:local-projection-on-integrals} is applied to all three terms with the same function $\chi$, charge sector and derivative order.
Thus the first difference removes the contribution of the continuum action, and the last term removes effects already accounted for by the known action coefficients.
We now express this residual as a sum of responses to local operators.
For each external correlator $A$, the response to one insertion of $\mathcal O_i^R$ is denoted by $\mathcal M_{Ai}$.
The unknown coefficients satisfy the linear equations
\begin{equation}
 \mathcal D_A^{[p,r]}=-\sum_i h_i^{[p,r]}\mathcal M_{Ai},
 \qquad
 \mathcal M_{Ai}=\mathcal P_{\rm loc}\!\left[
   \int d^2X\,
   \langle\mathcal A_A\mathcal O_i^R(X;q)\rangle_{G_{0,\epsilon},c}
   \right].
 \label{eq:free-correlator-matching}
\end{equation}
The same charge and derivative projections are used in $\mathcal D_A^{[p,r]}$ and $\mathcal M_{Ai}$.
The matrix $\mathcal M$ depends on the continuum correlation functions, the scales $q$ and $\epsilon$, the external positions, and the operator conventions.
All of these are held fixed, so neither $\mathcal M$ nor its inverse introduces any dependence on the lattice spacing $a$.
At the order being determined, $\mathcal M_{Ai}$ uses the Gaussian term in the renormalized operator insertion; contributions with already matched coefficients and further interaction insertions belong to the subtractions defining $\mathcal D_A^{[p,r]}$.
The minus sign follows by expanding the additional action in Eq.~\eqref{eq:symanzik}.
Choose enough elementary-field correlation functions and projections to distinguish the independent operators in the sector under consideration.
Here independence is assessed with the field convention fixed, without identifying operators by the equations of motion.
An invertible square set of these conditions gives
\begin{equation}
 h_i^{[p,r]}(a,q)=-\sum_A(\mathcal M^{-1})_{iA}\mathcal D_A^{[p,r]}.
 \label{eq:h-from-free-correlators}
\end{equation}
Thus the perturbative matching coefficients are obtained from free-boson correlation functions, their local subtractions and known lower-order coefficients.
A single two-point function need not distinguish all operators, so correlation functions with more field insertions may be needed.
The matching is performed at $\epsilon>0$, before taking the assumed infrared limit of the local coefficients.

\subsection{Example: determining two matching coefficients}
\label{sec:pairprojection}

We illustrate the matching procedure with two mass vertices of opposite charge.
Their local expansion can modify the quadratic derivative terms $:(\partial_x\phi)^2:_q$ and $:(\partial_\tau\phi)^2:_q$, whose coefficients we call $h_x$ and $h_\tau$.
We first calculate the connected correlator and select its two-derivative part.
We then subtract the continuum contribution and the effects of known coefficients, and solve for the remaining contributions to $h_x$ and $h_\tau$.
The final result, Eq.~\eqref{eq:pair-derivative-coefficients}, expresses their order-$m^2a^2$ contributions as subtracted integrals.

In this example, $\mathcal P_{\rm loc}$ uses the fixed sector $(k,d)=(0,2)$.
We abbreviate $\widehat V_{\pm,q,j}=\widehat V_{\pm1,q,j}$.
Take $\mathcal A_A=\phi(y_1)\phi(y_2)$ and two vertices at $X_\pm=X\pm r/2$, with $X$ away from $y_1,y_2$.
In theory $j=a,0$, define
\begin{equation}
 \begin{aligned}
 B_{j,\epsilon}(r)
 &=\langle\widehat V_{+,q,j}(X_+)\widehat V_{-,q,j}(X_-)\rangle_{G_{j,\epsilon},c},\\
 s_{j,b}(X)&=G_{j,\epsilon}(y_b-X),\qquad
 \Delta_rs_{j,b}=s_{j,b}(X+r/2)-s_{j,b}(X-r/2).
 \end{aligned}
 \label{eq:pair-correlator-definitions}
\end{equation}
For $j=0$ the extra subscript on $\widehat V$ is omitted in the earlier notation.
With each external field and each vertex a separate entry in the connected correlator, Wick contraction gives the exact identity
\begin{equation}
 \begin{aligned}
 F_{j,A}(X,r)
 &:=\langle\phi(y_1)\phi(y_2)
       \widehat V_{+,q,j}(X_+)\widehat V_{-,q,j}(X_-)\rangle_{G_{j,\epsilon},c}\\
 &=-\beta^2B_{j,\epsilon}(r)\,\Delta_rs_{j,1}\Delta_rs_{j,2}.
 \end{aligned}
 \label{eq:pair-connected-identity}
\end{equation}
Each external field can contract with either vertex, and their opposite charges give the differences $\Delta_rs_{j,b}$.
The connected subtraction is included in $B_{j,\epsilon}$; at separated points $B_{0,\epsilon}(r)\to|r|^{-2}$ as $\epsilon\to0$.
For the continuum term, the degree-two Taylor operation in Eq.~\eqref{eq:local-projection-definition} gives
\begin{equation}
 \Delta_rs_{0,b}=r_\mu\partial_\mu s_{0,b}(X)+O(|r|^3),
 \qquad
 [\Delta_rs_{0,1}\Delta_rs_{0,2}]_{d=2}
 =r_\mu r_\nu\partial_\mu s_{0,1}\partial_\nu s_{0,2}.
 \label{eq:pair-taylor}
\end{equation}
Indices $\mu,\nu$ in this example run over $x,\tau$ and are summed.
Define the local quadratic operators and their continuum responses by
\begin{equation}
 \mathcal O_{\mu\nu}=:\partial_\mu\phi\,\partial_\nu\phi:_q,
 \qquad
 \mathcal M_{A,\mu\nu}=\int d^2X\,
       \langle\phi(y_1)\phi(y_2)\mathcal O_{\mu\nu}(X)\rangle_{G_{0,\epsilon},c}.
 \label{eq:pair-operator-responses}
\end{equation}
Their Wick contractions are
\begin{equation}
 \langle\phi(y_1)\phi(y_2)\mathcal O_{\mu\nu}(X)\rangle_{G_{0,\epsilon},c}
 =\partial_\mu s_{0,1}\partial_\nu s_{0,2}
  +\partial_\nu s_{0,1}\partial_\mu s_{0,2}.
 \label{eq:pair-wick-response}
\end{equation}
Combining Eqs.~\eqref{eq:pair-taylor} and~\eqref{eq:pair-wick-response} using the integral projection in Eq.~\eqref{eq:local-projection-on-integrals} gives
\begin{equation}
 \begin{aligned}
 \mathcal P_{\rm loc}\!\left[\int d^2X\,d^2r\,F_{0,A}(X,r)\right]
 &=-\frac{\beta^2}{2}\sum_{\mu,\nu}I_{0,\mu\nu}\mathcal M_{A,\mu\nu},\\
 I_{j,\mu\nu}&:=\int_j d^2r\,\chi(qr)r_\mu r_\nu B_{j,\epsilon}(r).
 \end{aligned}
 \label{eq:pair-projected-moments}
\end{equation}
The factor $\chi$ enters through the projection and is included once in $I_{j,\mu\nu}$.
The same relative-coordinate moment on the lattice uses $a\sum_n\int dr_\tau$ with $r_x=na$.
Reflection symmetry gives $I_{j,x\tau}=0$.

To apply this result to the lattice--continuum difference, Fourier-transform the two external positions to momentum $\ell$.
At order $[p,r]=[2,0]$, $Z_a$ contributes its constant term one.
The two opposite-charge assignments in the two cosine insertions, together with the expansion factorial, give the factor $m^2/(4\pi^2)$ multiplying $F_{j,A}$.
Their unprojected contribution at $\epsilon>0$ is
\begin{equation}
 \mathcal C_{j,\rm pair}^{[2,0]}(\ell)
 =-\frac{m^2\beta^2}{2\pi^2}\,
       \widetilde G_{j,\epsilon}(\ell)^2
       \int_j d^2r\,B_{j,\epsilon}(r)
                  [1-\cos(\ell\cdot r)].
 \label{eq:pair-unprojected-fourier}
\end{equation}
The local projection inserts $\chi(qr)$ and replaces $1-\cos(\ell\cdot r)$ by $(\ell\cdot r)^2/2$, giving
\begin{equation}
 \mathcal C_{j,\rm pair}^{[2,0]}(\ell)\big|_{{\rm loc},d=2}
 =-\frac{m^2\beta^2}{4\pi^2}\,
       \widetilde G_{j,\epsilon}(\ell)^2
       \sum_{\alpha=x,\tau}I_{j,\alpha\alpha}\ell_\alpha^2.
 \label{eq:pair-projected-fourier}
\end{equation}
Here $d=2$ refers to the polynomial multiplying the two external propagators; those propagators are retained.
At fixed $\epsilon>0$, $\widetilde G_{a,\epsilon}(\ell)^2/\widetilde G_{0,\epsilon}(\ell)^2=1+O(|\ell|^4)$ near zero momentum.
Thus replacing the external lattice propagators by the continuum ones does not change this degree-two term.
The continuum responses are
\begin{equation}
 \mathcal M_{A,\alpha\alpha}(\ell)
 =2\widetilde G_{0,\epsilon}(\ell)^2\ell_\alpha^2,
 \qquad \alpha=x,\tau.
 \label{eq:pair-fourier-responses}
\end{equation}
For brevity write $\mathcal M_{A,\alpha}=\mathcal M_{A,\alpha\alpha}$ and define the already determined contribution $\kappa_{\alpha,\rm known}$ by
\begin{equation}
 \mathcal P_{\rm loc}\mathcal C_{A,\rm known}^{[2,0]}
 =-\sum_{\alpha=x,\tau}\kappa_{\alpha,\rm known}\mathcal M_{A,\alpha}.
 \label{eq:pair-known-projection}
\end{equation}
The coefficient $\kappa_{\alpha,\rm known}$ collects the part already generated by known effective-action insertions, including their contact terms.
It must be evaluated and subtracted, even in this two-vertex example.
In momentum space, we select charge zero, include the same relative-coordinate weight $\chi$, and extract the quadratic Taylor coefficient:
\begin{equation}
 \kappa_{\alpha,\rm known}
 =-\frac14\left.
 \frac{\partial^2}{\partial\ell_\alpha^2}
 \left[\widetilde G_{0,\epsilon}(\ell)^{-2}
       \mathcal C_{A,\rm known}^{[2,0]}(\ell)\right]_{k=0,\chi}
 \right|_{\ell=0}.
 \label{eq:pair-known-hessian}
\end{equation}
The division by $\widetilde G_{0,\epsilon}^2$ removes the two external propagators before the Taylor coefficient is taken.
In particular, the free dispersion insertion in Eq.~\eqref{eq:dispersion-correction} can act inside the vertex pair, even though its effect on the external propagators does not enter the degree-two term.
Using $\beta^2=4\pi$, the projected residual is therefore
\begin{equation}
 \mathcal D_A^{[2,0]}
 =-\sum_{\alpha=x,\tau}
 \left[\frac{m^2}{2\pi}(I_{a,\alpha\alpha}-I_{0,\alpha\alpha})
                   -\kappa_{\alpha,\rm known}\right]_{a^2}
 \mathcal M_{A,\alpha}.
 \label{eq:pair-matching-residual}
\end{equation}
The brackets retain the order-$a^2$ terms, including logarithms.
Choosing external momenta that distinguish $\ell_x^2$ from $\ell_\tau^2$ makes the two matching columns independent.
The inverse response matrix satisfies
\begin{equation}
 \sum_A(\mathcal M^{-1})_{\alpha A}\mathcal M_{A,\lambda}
 =\delta_{\alpha\lambda},\qquad \alpha,\lambda\in\{x,\tau\},
 \label{eq:pair-response-inversion}
\end{equation}
where $\delta_{\alpha\lambda}$ is the Kronecker delta.
Substitution in Eq.~\eqref{eq:h-from-free-correlators} therefore gives
\begin{equation}
 \left.h_\alpha^{[2,0]}\right|_{a^2}
 =\lim_{\epsilon\to0}\left[
       \frac{m^2}{2\pi}(I_{a,\alpha\alpha}-I_{0,\alpha\alpha})
       -\kappa_{\alpha,\rm known}\right]_{a^2},
 \qquad \mathcal O_\alpha=:(\partial_\alpha\phi)^2:_q.
 \label{eq:pair-derivative-coefficients}
\end{equation}
Equation~\eqref{eq:pair-derivative-coefficients} is the application of the general matching formula~\eqref{eq:h-from-free-correlators} to these two operators.
This formula specifies the required integrals but does not yet give numerical values or logarithmic coefficients.
Obtaining them requires evaluating the subtracted moments and taking the assumed infrared limit.
In particular, contributions to $(h_x+h_\tau)/2$ may cancel; counting derivatives alone cannot determine whether they do.

\subsection{Power counting inside the free correlators}
\label{sec:evenpowers}

Consider a group containing $p$ mass vertices and $r$ quadratic insertions in the expansion of Eq.~\eqref{eq:massless-correlator-expansion}.
The positions $y_{A,b}$ of the external fields remain fixed outside the group.
The group may be part of a term with additional interaction insertions; their positions are also held apart from the group when its points approach one another.
Resolve each mass vertex into charges $\sigma_s=\pm1$.
The resulting local operator has vertex charge
\begin{equation}
 k=\sum_{s=1}^{p}\sigma_s,\qquad |k|\le p,\qquad k\equiv p\pmod2.
 \label{eq:cluster-charge}
\end{equation}
The group need not be neutral.
In a higher-order term, mass vertices outside the group can carry the opposite charge, so a charged local term can contribute to a neutral correlation function with elementary external fields.
Charge neutrality of the complete set of vertex insertions therefore does not require $k=0$ for the group being replaced by a local operator.

The logarithmic Gaussian covariance gives the following leading operator-product coefficient inside these correlators, with $\xi_s=X_s-X$ the displacement of vertex $s$ from the common position $X$:
\begin{equation}
 \prod_{s=1}^{p}\widehat V_{\sigma_s,q}(X+\xi_s)
 \sim\left[\prod_{s<t}|\xi_s-\xi_t|^{2\sigma_s\sigma_t}\right]
       \left[\widehat V_{k,q}(X)+\text{derivative terms}\right].
 \label{eq:free-vertex-product}
\end{equation}
Under $\xi_s\mapsto\rho\xi_s$, the contraction factor scales as $\rho^{k^2-p}$ because $2\sum_{s<t}\sigma_s\sigma_t=k^2-p$.
The degree-$d$ Taylor operation in Eq.~\eqref{eq:local-projection-definition} supplies the factor $\xi^\alpha$, $|\alpha|=d$, in a projected moment $\langle T_k^R,\chi(q\xi)\xi^\alpha\rangle$ in Eq.~\eqref{eq:local-projection-distribution}.
Under the same rescaling this supplies a further $\rho^d$.
Each $Q_q$ has zero algebraic ultraviolet dimension: contractions involving its undifferentiated fields add logarithms, but no additional power of $\rho$.
The same holds for any undifferentiated fields left in the local operator.
Consequently, the coefficient of a charge-$k$ local term with $d$ derivatives scales as $\rho^{k^2+d-p}$, up to logarithms, in the local expansion of the free correlators.
The remaining correlation function contains that local operator at $X$, the elementary fields at $y_{A,b}$, and any interaction insertions outside the group; their separations are held fixed and do not contribute to this relative-coordinate power.

For $p+r\ge1$, one interaction point specifies the location $X$ of the local insertion and there are $p+r-1$ relative positions to integrate.
Putting these relative positions equal to $a$ times dimensionless variables supplies $a^{2(p+r-1)}$ from their measures, also for the spatial lattice sum and continuous time integral.
Combining this factor with the contraction and Taylor factors gives the algebraic power $\lambda=k^2+d+p+2r-2$ for a homogeneous short-distance contribution to the projected residual.
In a fixed $(k,d)$ sector, write its logarithmic dependence as $R_{A,p,r}(L)$.
Equation~\eqref{eq:h-from-free-correlators} then determines the corresponding polynomial $P_{i,p,r}(L)$ for the matching coefficient:
\begin{equation}
 \begin{aligned}
 \left.\mathcal D_A^{[p,r]}\right|_{\rm hom}
 &\sim m^p(g^2)^r a^\lambda R_{A,p,r}(L),\qquad L=\ln(aq),\\
 P_{i,p,r}(L)&=-\sum_A(\mathcal M^{-1})_{iA}R_{A,p,r}(L).
 \end{aligned}
 \label{eq:projected-power-counting}
\end{equation}
The restriction $|_{\rm hom}$ selects a contribution with the fixed power $\lambda$ under a common rescaling of the relative coordinates, allowing logarithmic factors.
Since $\mathcal M^{-1}$ is independent of $a$, it changes the coefficients of the logarithmic polynomial but introduces no further power of $a$.
The resulting homogeneous contribution to the matching coefficient is therefore
\begin{equation}
 \begin{aligned}
 h_i^{[p,r]}(a,q)&\sim
 m^p(g^2)^r
 \underbrace{a^{2(p+r-1)}}_{\text{relative measures}}\quad
 \underbrace{a^{k^2+d-p}}_{\text{local operator product}}
 P_{i,p,r}(\ln(aq))\\
 &=m^p(g^2)^r a^{k^2+d+p+2r-2}P_{i,p,r}(\ln(aq)).
 \end{aligned}
 \label{eq:local-power}
\end{equation}
Numerical constants and the dependence on fixed operator conventions are contained in $P_{i,p,r}$.
The sum over external correlation functions is therefore contained in $P_{i,p,r}$; it can cancel a logarithmic term or the whole contribution.
Integrations between $a$ and $q^{-1}$ and the local subtractions can produce further logarithms.
As in the local perturbative expansion assumed above, we take their dependence to be a finite polynomial at each fixed order.
For the explicit pair projection in Sec.~\ref{sec:pairprojection}, $(p,r,k,d)=(2,0,0,2)$ gives the possible scale $m^2a^2$ times logarithms for the subtracted derivative coefficients.
Three mass vertices with total charge one and no derivatives have $(p,r,k,d)=(3,0,1,0)$ and give the possible scale $m^3a^2$ for the fundamental vertex coefficient.

Independent reflections of space and Euclidean time require an even number of each type of derivative in a bulk action term, so $d$ is even.
Together with Eq.~\eqref{eq:cluster-charge}, this implies
\begin{equation}
 k^2+d+p+2r-2\equiv k+p\equiv0\pmod2.
 \label{eq:evenpowers}
\end{equation}
The sector without interaction insertions is given by the free-propagator expansion, Eq.~\eqref{eq:free-matching}, which also contains only even powers.
Repeated lower-order insertions and their subtractions preserve this parity.
After the continuum parameters are matched and vacuum terms removed, the first possible positive power is therefore $a^2$ within the assumed local expansion.
This count determines allowed algebraic powers.

\FloatBarrier
\bibliographystyle{utphys}
\bibliography{refs}

\end{document}